\documentclass[runningheads]{llncs}

\usepackage{amssymb}
\usepackage{graphicx}
\usepackage{amsmath}
\usepackage{latexsym,amssymb,color}
\usepackage{ifthen,graphics}
\usepackage{xspace}

\newcommand{\Right}{$\mathsf{right}$}
\newcommand{\Down}{$\mathsf{down}$}
\newcommand{\Stop}{$\mathsf{stop}$}
\newcommand{\Set}{$\mathsf{set}$}
\newcommand{\Get}{$\mathsf{get}$}

\newcommand{\TLA}{TLA\kern-1pt$^+$\xspace}

\newcommand{\m}[1]{\ensuremath{\mathcal{#1}}\xspace}
\newcommand{\ID}{\mbox{\rm ID}}
\newcommand{\id}{\mbox{\rm id}}

\renewcommand{\dim}{\mbox{\sl dim}}

\newcommand {\CAS} {{\sf Compare\&Swap}\xspace}
\newcommand {\TAS} {{\sf Test\&Set}\xspace}
\newcommand {\FAA} {{\sf Fetch\&Add}\xspace}
\newcommand {\SWAP} {{\sf Swap}\xspace}
\newcommand {\R} {{\sf Read}\xspace}

\newcommand {\W} {{\sf Write}\xspace}

\newcounter{linecounter}
\newcommand{\linenumbering}{\ifthenelse{\value{linecounter}<10}{(0\arabic{linecounter})}{(\arabic{linecounter})}}
\renewcommand{\line}[1]{\refstepcounter{linecounter}\label{#1}\linenumbering}
\newcommand{\resetline}[1]{\setcounter{linecounter}{0}#1}
\renewcommand{\thelinecounter}{\ifnum \value{linecounter} > 9\else 0\fi \arabic{linecounter}}

\begin{document}

\title{Tasks in Modular Proofs\\ of Concurrent Algorithms}

\author{Armando Casta\~neda\inst{1} \and
Aur\'elie Hurault\inst{2} \and 
Philippe Qu\'einnec\inst{2} \and \\
Matthieu Roy\inst{1,3}}
\authorrunning{A. Casta\~neda et al.}
\titlerunning{Tasks in Modular Proofs of Concurrent Algorithms}

\institute{Instituto de Matem\'aticas, UNAM, M\'exico \and
IRIT -- Universit\'e de Toulouse, France \and
LAAS, CNRS, Toulouse, France\\
\email{armando.castaneda@im.unam.mx, \{hurault,queinnec\}@enseeiht.fr, roy@laas.fr}}
\maketitle              
\begin{abstract}
\noindent Proving correctness of distributed or concurrent algorithms is a mind-challenging
and complex process. Slight errors in the reasoning are difficult to find, calling for
computer-checked proof systems.
In order to build computer-checked proofs with usual tools, such as
Coq or \TLA, having sequential specifications
of all base objects that are used as building
blocks in a given algorithm is a requisite to provide a modular
proof built by composition. Alas, many concurrent objects do not
have a sequential specification.

This article describes a systematic method to transform any {\em task}, a specification method 
that captures concurrent one-shot distributed problems, into
a sequential specification involving two calls, {\Set} and {\Get}.
This transformation allows system designers to {\em compose} proofs,
thus providing a framework for modular computer-checked proofs
of algorithms designed using tasks and sequential objects as building blocks.
The  Moir\&Anderson implementation of \emph{renaming} using \emph{splitters} is an iconic example
of such algorithms designed by composition. 

~\\
\textbf{Keywords:} Formal methods $\cdot$ Verification $\cdot$ Concurrent algorithms $\cdot$ Renaming.
\end{abstract}

\section{Introduction}

Fault-tolerant distributed and concurrent algorithms are extensively used in critical systems
that require strict guarantees of correctness~\cite{IEC-61508};
consequently, verifying such algorithms is becoming more important nowadays.
Yet, proving distributed and concurrent algorithms is a difficult and error-prone task,
due to the complex interleavings that may occur in an execution. 
Therefore, it is crucial to develop frameworks that help assessing the correctness of such systems.

%

A major breakthrough in the direction of systematic proofs of concurrent algorithms
is the notion of 
\emph{atomic} or \emph{linearizable} objects~\cite{HW90}:
a linearizable object behaves as if it is accessed sequentially, even in presence of concurrent invocations, the canonical example being the atomic register. 
Atomicity lets us model a concurrent algorithm as a transition system in which each transition
corresponds to an atomic step performed by a process on a base object.  Human beings
naturally reason on sequences of events happening
one after the other; concurrency and interleavings seem to be more difficult to 
deal with.

However, it is well understood now that several natural one-shot base objects used in concurrent algorithms cannot be expressed
as sequential objects~\cite{CRR18,HRV15,N94} providing a single operation. 
%

An iconic example is the \emph{splitter} abstraction~\cite{MoirA95}, which is the basis of the classical Moir\&Ander\-son renaming algorithm~\cite{MoirA95}.
Intuitively, a splitter is a concurrent one-shot problem that splits calling processes as follows: whenever $p$ processes access 
a splitter, at most one process obtains {\Stop}, at most $p-1$ obtain {\Right} and at most $p-1$ obtain {\Down}.
Moir\&Anderson renaming algorithm 
uses splitters arranged in a half grid to scatter processes and provide new names to processes.
It is worth to mention that,  since its introduction almost thirty years ago, 
the \emph{renaming} problem~\cite{ABDPR90} has become a paradigm for studying symmetry-breaking in concurrent systems (see, for example,~\cite{A15,CRR11}).

A second example is the \emph{exchanger} object provided in Java, which has been used for implementing efficient linearizable elimination stacks~\cite{HRV15,SLS09,ST97}.
Roughly speaking, an exchanger is a meeting point where pairs of processes can exchange values, with the constraint that an exchange 
can happen only if the two processes run concurrently. 

Splitters and exchangers are instances of one-shot concurrent objects known in the literature as {\em tasks}. 
Tasks have played a fundamental role 
in understanding the computability power of several models, providing 
a topological view of concurrent and distributed computing~\cite{HKRbook}. 
Intuitively, a task is an object providing a single one-shot operation, 
formally specified through an input domain, an output domain and an input/output relation describing 
the valid output configurations when a set of processes run concurrently, starting from a given input configuration.
Tasks can be equivalently specified by mappings between topological objects: an input simplicial complex (i.e., a discretization of a continuous topological space)
modeling all possible input assignments,  an output simplicial complex modeling all possible output assignments, and a carrier map relating inputs and outputs.

\paragraph{Contributions.}
Our main contribution is a generic transformation of any task $T$ (with a single operation) into
a sequential object
$S$ providing two operations, {\Set} and {\Get}.
The behavior of $S$ ``mimics'' the one of $T$ by splitting each invocation 
of a process to $T$ into two 
invocations to $S$, first {\Set} and then {\Get}. Intuitively, the {\Set} operation 
records the processes
that are participating to the execution of the task. A process actually calls the task and obtains a return value by invoking {\Get}.
Each of the operations is atomic; however,  {\Set} and {\Get} invocations of a given process may
be interleaved with similar invocations from other processes.

We show that these two operations are sufficient for any task, no matter how complicated it may be; since a task is a mapping between 
simplicial complexes, it can specify very complex concurrent behaviors, 
sometimes with obscure associated operational semantics.

A main benefit of our transformation is that one can replace an object solving a task $T$ by its associated sequential object $S$, and reason as if all steps
happen sequentially. This allows us to obtain simpler models of concurrent algorithms
using solutions to tasks and sequential objects as building blocks, leading to modular correctness proofs.
Concretely, we can obtain a simple transition system of Moir\&Anderson renaming algorithm,
which helps to reason about it. In a companion paper~\cite{HQ19}, our model is used to derive a full and modular \TLA proof of the algorithm,
the first available \TLA proof of it.




In Section~\ref{sec:ma-renaming}, we explain the ideas in Moir\&Anderson renaming algorithm that motivated our general transformation,
which is presented in Section~\ref{sec-dealing-tasks}. 
Due to lack of space, some basic definitions, proofs and detailed constructions are omitted. They can be found in the extended version~\cite{SSS19long}.

\section{Verifying Moir\&Anderson Renaming}
\label{sec:ma-renaming}

We consider a concurrent system with $n$ asynchronous processes, meaning that each process can experience arbitrarily long delays during an execution. Moreover, processes may crash at any time, i.e., permanently stopping taking steps.
Each process is associated with a unique $\ID \in  \mathbb{N}$. The processes can access \emph{base}
objects like simple atomic read/write registers or more complex objects.


The original Moir\&Anderson renaming algorithm~\cite{MoirA95} is designed and explained with splitters. Their seminal work first introduces the splitter algorithm based on atomic read/write registers and discusses its properties.
Then, they describe a renaming algorithm that uses a grid of splitters. The
actual implementation inlines splitters into the code of the renaming algorithm,
and their proof is performed on the resulting program that uses solely read/write registers
as base objects.

\paragraph{The Splitter Abstraction.} A \emph{splitter}~\cite{MoirA95}  is a one-shot concurrent task in which
each process starts with its unique $\ID \in  \mathbb{N}$ and has to return a value satisfying the following  properties:
(1) ${\sf Validity}$. 
 The returned value is \Right, \Down \,  or \Stop. 
(2) ${\sf Splitting}$. If $p \geq 1$ processes participate in an execution of the splitter, then
at most $p-1$ processes  obtain  the value \Right, at most $p-1$  processes obtain  the value \Down, 
at most one  process obtains the value \Stop. 
(3) ${\sf Termination}$. Every correct process (which doesn't crash) returns a value.

Notice that if a process runs solo, i.e., $p=1$, it must obtain \Stop, since the $\mathsf{splitting}$ property  holds for any $p\geq 1$.

\begin{figure}[!ht] 
\begin{minipage}{.5\linewidth}
\centering{
\fbox{
\begin{minipage}[t]{150mm}
\scriptsize
\renewcommand{\baselinestretch}{2.5} 
\resetline
\begin{tabbing}
aaaaa\=aaaaa\=aaaaaaaaaaaaa\=aaa\=\kill 

{\bf initially} $\mathit{CLOSED} = false$ \\

{\bf operation}  
     $\mathsf{splitter}()$:  \\

\line{SPLIT01}
\>  $\mathit{LAST} \leftarrow my\_ID$;\\

\line{SPLIT02}
\>  {\bf if} \=  $(\mathit{CLOSED})$ \\

\line{SPLIT03} 
 
\> \>  {\bf then}  \= $\mathsf{return}(\mathsf{right})$ \\

\line{SPLIT04} 

\> \> {\bf else}  \>
 
$\mathit{CLOSED} \leftarrow \mathit{true}$;\\

\line{SPLIT05}  
\> \>  \> {\bf if} \=  ($\mathit{LAST} =  my\_ID$)\\

\line{SPLIT06}  \> \>  \> \= 

    {\bf then}  \= ${\mathsf{return}}(\mathsf{stop})$ \\

\line{SPLIT07}  \> \>  \> \>

   {\bf else} \>   ${\mathsf{return}}(\mathsf{down})$ \\

\line{SPLIT08}  
\> \>  \> {\bf end if}\\

\line{SPLIT09}
\>  {\bf end if}.

\end{tabbing}
\normalsize
\end{minipage}
}
\caption{Implementation of a Splitter~\cite{MoirA95}.}
\label{algo:algo-splitter}
}
 
\end{minipage} \hfill   
\begin{minipage}{.5\linewidth}  
\centering{
\includegraphics[width=.55\linewidth]{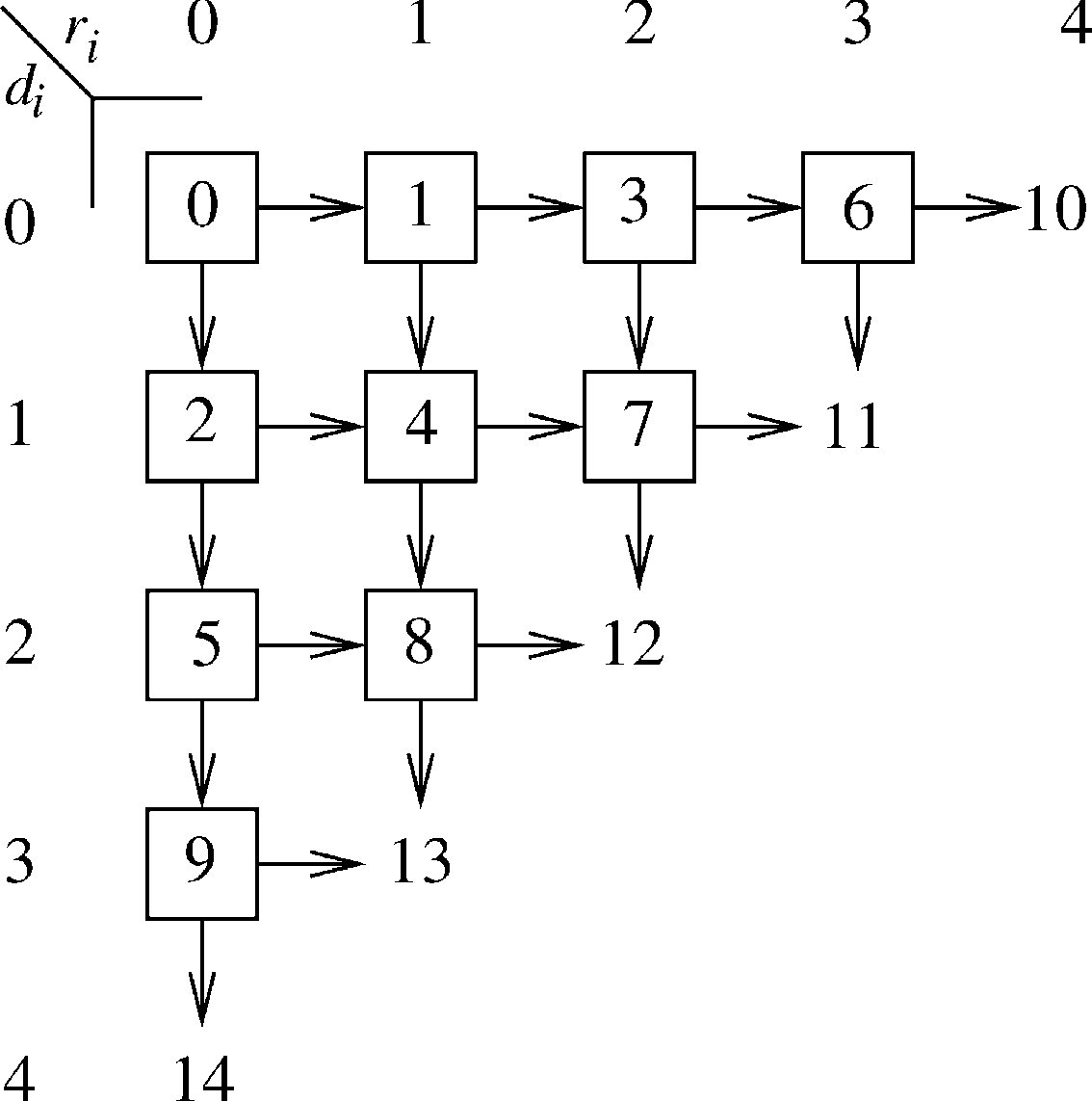}}
\caption{Renaming using Splitters.}
\label{fig:splitter-grid}
\end{minipage}
\end{figure}

%
%

Figure~\ref{algo:algo-splitter} contains the simple and elegant 
splitter implementation based on atomic read/write registers from~\cite{MoirA95} (register
names  have been changed for clarity).
After carefully analyzing the code, the reader can convince herself that the algorithm 
described in Figure~\ref{algo:algo-splitter} implements the splitter specification. 
The fact that the implementation is based on \emph{atomic} registers
allows us to obtain a transition system of it in which each transition corresponds to an atomic operation on an object. The benefit of this modelization is that  
every execution of the implementation is simply described as a sequence of steps, 
as concurrent and distributed systems are usually modeled (see, for example,~\cite{HS08,R13})).
Although the splitter implementation is very short and simple,
its \TLA proof is long and rather complex ---particularly when considering that it uses 
a boolean register and a plain register only--- (see~\cite{HQ19} for details). 


%
%

\paragraph{The Renaming Problem.} 
In the \emph{$M$-renaming} task~\cite{ABDPR90}, 
each process starts with its unique $\ID \in  \mathbb{N}$, and processes are required to 
return an output name satisfying the following properties:
(1) ${\sf Validity}$. The output name of a process belongs to $[1, \hdots, M]$.
(2) ${\sf Uniqueness}$. No two processes obtain the same output name.
(3) ${\sf Termination}$. Every correct process returns an output name.


Let $p$ be the number of  processes that participate in a given renaming instance. 
A renaming implementation is \emph{adaptive} if 
the  size $M$   of  the  new name  space  depends only on  $p$, 
the number of participating processes.  We have then $M=f(p)$ 
where  $f(p)$ is a function on $p$ such that  $f(1)=1$ and, 
for $2\leq p \leq n$,  $p-1 \leq f(p-1)\leq f(p)$.  



\paragraph{Moir\&Anderson Splitter-Based Renaming Algorithm.}
Moir and Anderson propose in~\cite{MoirA95} a read/write renaming algorithm designed
using the splitter abstraction. 
The algorithm is conceptually simple: for up to $n$ processes, a set of $n(n+1)/2$ splitters are placed in a half-grid, each with a unique name, as shown in Figure~\ref{fig:splitter-grid} for $n=5$.
Each process starts invoking the splitter at the top-left corner, following the directions
obtained at each splitter. When a splitter invocation returns stop, the process returns the name associated with the splitter. 
We use here an adaptive version of their algorithm
that allows $p$ participating processes to rename in at most $p(p+1)/2$ names; the original solution in~\cite{MoirA95} is non-adaptive and the only difference 
is the labelling of the splitters in the grid.

\paragraph{Splitters as Sequential Objects?}
Although Moir\&Anderson renaming algorithm is easily described in a modular way,
the actual program is not modular as each splitter in the conceptual grid is replaced by an independent copy of the splitter implementation of Figure~\ref{algo:algo-splitter}.
Thus, the correctness proof in~\cite{MoirA95} deals with the possible interleavings that can occur, considering 
all read/write splitter implementations in the grid. 

In the light of the simple splitter based conceptual description, we would like to have a transition system
describing the algorithm based on splitters as building blocks,
in which each step corresponds to a splitter invocation.
Such a description would be very beneficial as it would allow us to obtain a modular correctness proof showing that the algorithm is correct as long as the building blocks are splitters, 
hence the correctness is independent of any particular splitter implementation.

\begin{figure}[tb]
\centering{
\begin{minipage}[t]{80mm}
\footnotesize
\begin{tabbing}
aa\=aaaa\=aaaa\=aaaa\=\kill 

{\bf State:} Sets $Participants, Stop, Down, Right$\\
all sets are initialized to $\emptyset$\\ \\

{\bf Function} {\Set}($\id$)\\

\> {\bf Pre-condition:} $\id \notin Participants$\\

\> {\bf Post-condition:} $Participants' \leftarrow Participants \cup \{ \id \}$\\

\> {\bf Output:} $\sf void$\\

{\bf endFunction}\\ \\

{\bf Function} {\Get}($\id$)\\

\> {\bf Pre-condition:} $\id \in Participants \wedge \id \notin Stop, Down, Right$\\

\> {\bf Post-condition:}\\ 

\> \> $D \leftarrow \emptyset$\\

\> \> {\bf if $|Stop| = 0$ then} $D \leftarrow D \cup \{\mathsf{stop}\}$\\ 

\> \> {\bf if $|Down| < |Participants|-1$ then} $D \leftarrow D \cup \{\mathsf{down}\}$\\ 

\> \> {\bf if $|Right| < |Participants|-1$ then} $D \leftarrow D \cup \{\mathsf{right}\}$\\ 

\> \> Let $dec$ be any value in $D$\\

\> \> {\bf if $dec = $ {\Stop} then} $Stop \leftarrow Stop \cup \{\id\}$\\ 

\> \> {\bf if $dec = $ {\Down} then} $Down \leftarrow Down \cup \{\id\}$\\ 

\> \> {\bf if $dec = $ {\Right} then} $Right \leftarrow Right \cup \{\id\}$\\ 

\> {\bf Output:} $dec$\\

{\bf endFunction}

\end{tabbing}
\footnotesize
\end{minipage}
\caption{An \textit{ad hoc} specification of the Splitter.}
\label{fig-splitter-spec}
}
\end{figure}

As it is formally proved in Section~\ref{sec-dealing-tasks}, it is impossible to obtain such a transition system.
The obstacle is that a splitter is inherently concurrent and cannot be specified as a sequential object with a single operation.
The intuition of the impossibility is the following.
By contradiction, suppose that there is a sequential object corresponding to a splitter. Since the object is sequential,
in every execution, the object behaves as if it is accessed sequentially (even in presence of concurrent invocation).
Then, there is always a process that invokes the splitter object first, which, as noted above, must obtain {\Stop}.
The rest of the processes can obtain either {\Down} or {\Right}, without any restriction (the value obtained by the first process
precludes that all obtain {\Right} or all {\Down}). However, such an object is allowing strictly fewer behaviors: 
in the original splitter definition it is perfectly possible that all processes run concurrently and half of them obtain {\Right}
and the other half obtain {\Down}, while none obtains {\Stop}.


\paragraph{The splitter task as a sequential object.}
One can circumvent the impossibility described above by splitting the single method provided by a splitter into two (atomic) operations of
a sequential object. Figure~\ref{fig-splitter-spec} presents a sequential specification of a splitter with two operations, {\Set} and {\Get}, using a standard pre/post-condition specification style.
Each process invoking the splitter, first invokes {\Set} and then {\Get} (always in that order).
The idea is that the {\Set} operation first records in the state of the object the processes that are participating in the splitter, so far, and then
the {\Get} operation nondeterministically produces an output to a process, considering the rules of the splitter.
In Section~\ref{sec-dealing-tasks}, we formally prove that this sequential object indeed models the splitter defined above.
\paragraph{Proving Moir\&Anderson Renaming with Splitters as Base Sequential Objects.}

Using the sequential specification of a splitter in Figure~\ref{fig-splitter-spec}, we can easily obtain a \emph{generic}
description of the original Moir\&Anderson splitter-based algorithm: each renaming object is replaced with 
an equivalent sequential version of it, and every process accessing a renaming object asynchronously invokes first {\Set}
and then {\Get}, which returns a direction to the process.
The resulting algorithm does not rely on any particular splitter implementation, and uses only atomic objects,
which allows us to obtain a transition system of it.
This is the algorithm that is verified in \TLA in~\cite{HQ19}. 
The equivalence between the concurrent renaming specification and the sequential {\Set}/{\Get} specification
imply that the proof in~\cite{HQ19} also proves for the original Moir\&Anderson splitter-based algorithm.

\section{Dealing with Tasks without Sequential Specification}
\label{sec-dealing-tasks}

In this section, we show that the transformation in Section~\ref{sec:ma-renaming} of the splitter task into a sequential
object with two operations, {\Get} and {\Set}, is not a trick but rather a general methodology to deal with tasks without a sequential specification.
Our {\Get}/{\Set} solution proposed here is reminiscent to the \emph{request-follow-up} transformation in~\cite{SS04}
that allows to transform a \emph{partial} method of a sequential object (e.g. a queue with a blocking dequeue method when the queue is empty) 
into two \emph{total} methods:
a total request method registering that a process wants to obtain an output, and a total follow-up method 
obtaining the output value, or \emph{false} if the conditions for obtaining a value are not yet satisfied
(the process invokes the follow-up method until it gets an output).
We stress that the \emph{request-follow-up} transformation~\cite{SS04} considers only objects with a sequential specification
and is not shown to be general as it is only used for queues and stacks.

\paragraph{Model of Computation in Detail.}
We consider a standard concurrent system
with $n$ \emph{asynchronous} processes, $p_1, \hdots, p_n$,
which may \emph{crash} at any time during an execution of the system, i.e., 
stopping taking steps (for more detail see for example~\cite{HS08,R13}).
{Processes communicate with each other by invoking operations
{on} shared, concurrent \emph{base objects}.
A base object can provide $\R/\W$ operations (also called \emph{register}), 
more powerful operations, such as  $\TAS, \FAA, \SWAP \textrm{~or~} \CAS$, 
or solve a concurrent distributed problem, for example, $\mathsf{Splitter}$, $\mathsf{Renaming}$ or $\mathsf{Set\_Agreement}$.

Each process follows a local state machines $A_1, \hdots, A_n$,
where $A_i$ specifies which operations on base objects $p_i$ executes
in order to return a response when it invokes a high-level operation (e.g. $\mathsf{push}$ or $\mathsf{pop}$ operations).
Each of these base-objects operation invocations is a \emph{step}.
An \emph{execution} is a possibly infinite sequence of steps
and invocations and responses of high-level operations,
with the following properties:
\begin{enumerate}

\item Each process first invokes a high-level operation, and only when it has a corresponding response,
it can invoke another high-level operation, i.e., executions are \emph{well-formed}.

\item For any invocation $inv(\langle \mathsf{opType}, p_i, input \rangle)$ of a process $p_i$,
the steps of $p_i$ between that invocation and {its corresponding response (if there is one),
are steps} that are specified by $A_i$ when $p_i$
invokes the high-level operation $\langle \mathsf{opType}, p_i, input \rangle$.
\end{enumerate}

A high-level operation in an execution is \emph{complete} if
both its invocation and response appear in the execution.
An operation is \emph{pending}
if only its invocation appears in the execution.
{A process is \emph{correct} in an execution if it takes infinitely many steps.}

\paragraph{Sequential Specifications.}

A central paradigm for specifying distributed problems is that of a shared object $X$ that processes may access
concurrently~\cite{HS08,R13}, but the object is defined in terms
of a \emph{sequential specification}, i.e., an automaton describing the
outputs the object produces when it is accessed sequentially. 
Alternatively, the specification can be described as (possibly infinite) prefix-closed set, $SSpec(X)$,
with all the sequential executions allowed by $X$.

Once we have a sequential specification, there are various ways of defining what it means
for an execution to \emph{satisfy} an object, namely, that it respects the sequential
specification. \emph{Linearizability}~\cite{HW90} is the standard notion used to identify
correct executions of implementations of sequential objects.
Intuitively, an execution is linearizable if its operations can be
ordered sequentially, without reordering {non-overlapping} operations,
so that their responses satisfy the specification of the implemented object.
To formalize this notion we define a partial order on
the completed operations of an execution $E$:
$\mathsf{op} <_E \mathsf{op}'$ if and only if the response of $\mathsf{op}$ precedes
the invocation of $\mathsf{op}'$ in $E$.
Two operations are \emph{concurrent} if they are incomparable by $<_E$.
The execution is \emph{sequential} if $<_E$ is a total order.

An execution $E$ is \emph{linearizable} with respect to $X$ if there is
a sequential execution $S$ of $X$ (i.e., $S \in SSpec(X)$) such that:
(1) $S$ contains every completed operation of $E$
and might contain some pending operations.
Inputs and outputs of invocations and responses in $S$ agree with inputs and outputs in~$E$,
and (2) for every two completed operations $\mathsf{op}$ and $\mathsf{op}'$ in $E$,
if $\mathsf{op} <_E \mathsf{op}'$, then $\mathsf{op}$ appears before $\mathsf{op}'$ in $S$.

Using the linearizability correctness criteria for sequential objects, we can define the set of \emph{valid} executions for $X$, denoted $VE(X)$, as 
the set containing every execution $E$ that consists of invocations and responses and is linearizable w.r.t.  $X$.
$VE(X)$ contains the behavior one might expect from any \emph{building-block} implementation of $X$, e.g., any algorithm that implements $X$.

\paragraph{Tasks.}

A task is the basic distributed equivalent of a function in sequential computing, defined by a
set of inputs to the processes and for each (distributed) input to the
processes, a set of legal (distributed) outputs of the processes,
e.g.,~\cite{HKRbook}.  

In an algorithm designed to solve a task, each
process starts with a private input value and has to eventually decide
irrevocably on an output value. 
A process $p_i$ is initially not
aware of the inputs of other processes.  Consider an execution where
only a subset of $k \leq n$ processes participate; the others crash without
taking any steps. A  set of pairs
$\sigma=\{(\id_1,x_1),\dots,(\id_k,x_k)\}$ is used to denote the input values,
or output values, in the execution, where  $x_i$ 
denotes the value of the process with identity $\id_i$, either an input
value or an output value. 
A set $\sigma$ as above is called a \emph{simplex}, and if the values
are input values, it is an \emph{input simplex}, if they are output
values, it is an \emph{output simplex}.  The elements of $\sigma$ are
called \emph{vertices}, and any subset of $\sigma$ is a \emph{face} of it.
An \emph{input vertex} $v=(\id_i,x_i)$
represents the initial state of process $\id_i$, while an \emph{output
  vertex} represents its decision.  The \emph{dimension} of a simplex
$\sigma$, denoted $\dim(\sigma)$, is $|\sigma|-1$, and it is
\emph{full} if it contains $n$ vertices, one for each process.  
A \emph{complex} $\m{K}$ is a set of simplexes (i.e. a set of sets) closed under containment.
The dimension of $\m{K}$ is the largest dimension of its
simplexes, and $\m{K}$ is \emph{pure} of dimension $k$ if each of its
simplexes is a \emph{face} of a $k$-dimensional simplex. In distributed
computing, the simplexes (and complexes) are often \emph{chromatic}:
vertices of a simplex are labeled with a distinct process identities.  
The set of processes identities in an input or output
simplex $\sigma$ is denoted $\ID(\sigma)$.

\begin{figure}[tb]
\centerline{\includegraphics[width=10.5cm]{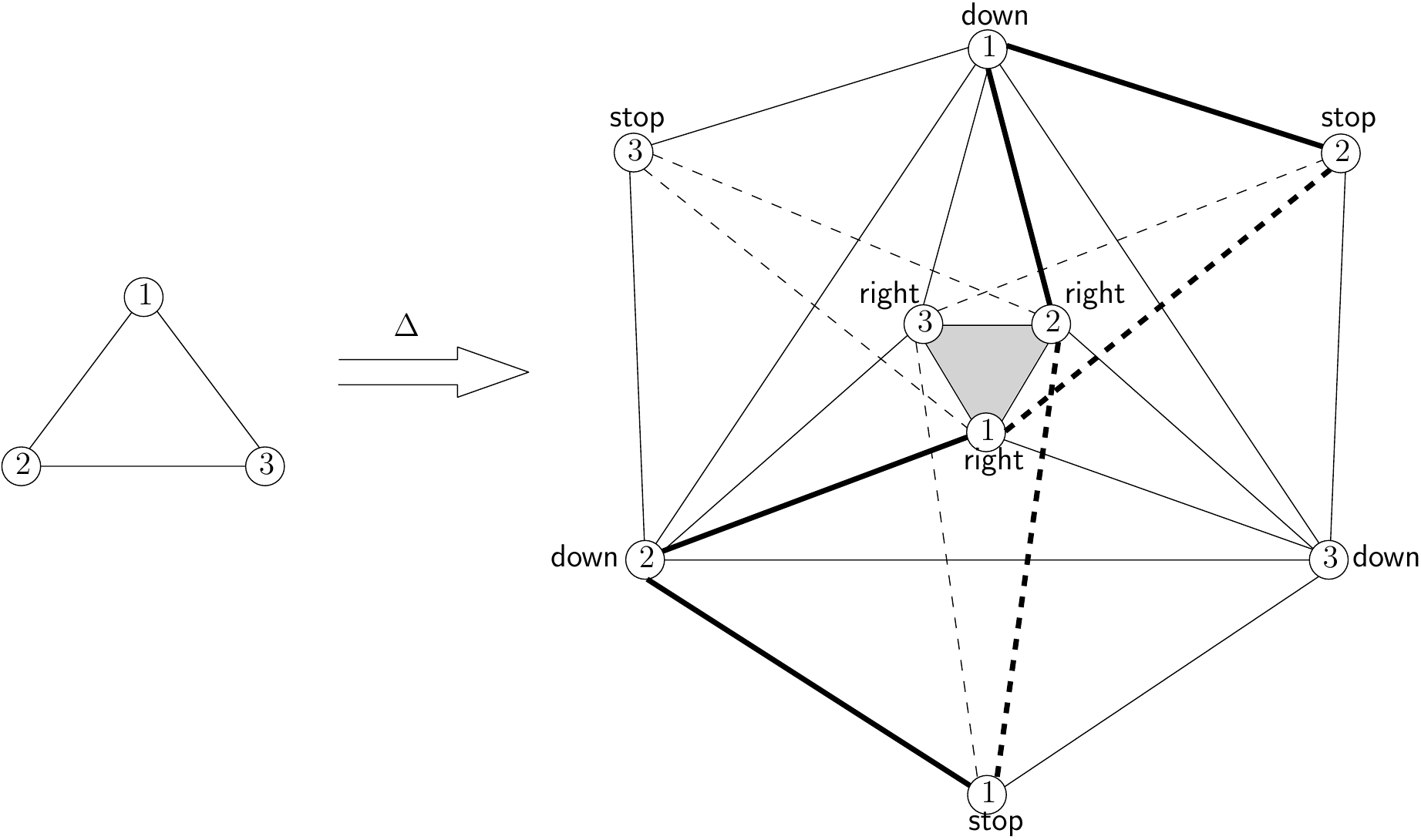}}
\caption{The Splitter Task for Three Processes.}
\label{fig-splitter}
\end{figure}

A \emph{task} $T$ for $n$ processes is a triple $(\m{I},\m{O},\Delta)$
where $\m{I}$ and $\m{O}$ are pure chromatic $(n-1)$-dimensional
complexes, and $\Delta$ maps each simplex $\sigma$ from $\m{I}$ to a
subcomplex $\Delta(\sigma)$ of $\m{O}$, satisfying:
(1) $\Delta(\sigma)$ is pure of dimension $\dim(\sigma)$,
(2) for every $\tau$ in $\Delta(\sigma)$ of dimension
  $\dim(\sigma)$, $\ID(\tau) = \ID(\sigma)$,
and (3) if $\sigma,\sigma'$ are two simplexes in $\m{I}$ with
$\sigma'\subset \sigma$ then $\Delta(\sigma')\subset\Delta(\sigma)$.
A task is a very compact way of specifying a distributed problem, and indeed
typically it is hard to understand what exactly is the problem being
specified.  Intuitively, $\Delta$ specifies, for every simplex
$\sigma\in\m{I}$, the valid outputs $\Delta(\sigma)$ for the processes
in $\ID(\sigma)$ assuming they run to completion, and the other
processes crash initially, and do not take any steps.

As an example consider the \emph{splitter} task~\cite{MoirA95}.
Figure~\ref{fig-splitter} shows a graphic description of the splitter task for three processes with IDs 1, 2 and 3.
The input complex, shown at the left, consists of a triangle and all its faces. The output complex, at the right, 
contains all possible valid output simplexes (the triangle with all {\Right} outputs is not in the complex).
The $\Delta$ function maps the input vertex with $\ID$ $1$ to the output vertex $(1,\mathsf{stop})$, the input edge 
with $\ID$s 1 and 2 to the complex with the bold edges in the output complex, and the input triangle is mapped to
the whole output complex. The rest of $\Delta$ is defined symmetrically.

Let $E$ be an execution where each process invokes a task $\langle \m{I},
\m{O}, \Delta \rangle$ once.  Then, $\sigma_{E}$ is the input simplex
defined as follows: $(\id_i,x_i)$ is in $\sigma_{E}$ iff in $E$ there
is an invocation of ${\sf task}(x_i)$ by process $\id_i$.  The output
simplex $\tau_{E}$ is defined similarly: $(\id_i,y_i)$ is in
$\tau_{E}$ iff there is a response $y_i$ to a process $\id_i$ in $E$.
We say that $E$ \emph{satisfies} $(\m{I}, \m{O}, \Delta)$ if for
every prefix $E'$ of $E$, it holds that $\tau_{E'} \in
\Delta(\sigma_{E'})$.

Using the satisfiability notion of tasks we can now consider the set of valid executions, $VE(T)$,
for a given task $T = (\m{I},\m{O},\Delta)$: 
the set containing every execution $E$ that has only invocations and responses and satisfies $T$.
Arguably, the set $VE(T)$ contains the behavior one might expect from a \emph{building-block} (e.g. an algorithm) that implements $T$. 

\paragraph{Modeling Tasks as Sequential Objects.}

Intuitively, tasks and sequential specifications are inherently different paradigms for specifying distributed problems:
while a task specifies what a set of processes might output when running concurrently, a sequential specification
specifies the behavior of a concurrent object when accessed sequentially (and linearizability tells when a 
concurrent execution ``behaves'' like a sequential execution of the object).
A natural question is if any task can be modeled as a sequential object with a single operation, namely, the object defines the same set of valid executions. 
A well-known example for which this is 
possible is the consensus distributed coordination problem that can be equivalently defined as a task or as a sequential 
object (see for example~\cite{HS08} where it is defined as an object\footnote{Sometimes, for clarity or efficiency, the object is defined with two operations 
(in the style of the Theorem~\ref{theo-two-ops}); however, consensus can be equivalently defined with one operation.} 
and~\cite{HKRbook} where it is defined as a task). 

\begin{lemma}
\label{lemma-no-seq-spec}
Consider the splitter task $T_\mathsf{spl} = (\m{I}_\mathsf{spl},\m{O}_\mathsf{spl},\Delta_\mathsf{spl})$.
There is no sequential object $X_\mathsf{spl}$ with a single operation satisfying
$VE(T_\mathsf{spl}) = VE(X_\mathsf{spl})$.
\end{lemma}
%

In a very similar way, one can prove that 
the following known tasks cannot be specified as sequential objects
with a single operation: \emph{exchanger}~\cite{HSY10,ST97}, \emph{adaptive renaming}~\cite{ABDPR90},
\emph{set agreement}~\cite{C93}, \emph{immediate snapshot}~\cite{BG93},
\emph{adopt-commit}~\cite{BGLR,G98} and \emph{conflict detection}~\cite{AE14}.\footnote{There are non-deterministic
sequential specifications of these tasks with \emph{unavoidable} and \emph{pathological} executions in which some operations \emph{guess} the inputs of future operations.
See \cite[Section 2]{CRR18} for a detailed discussion.}

To circumvent the impossibility result in Lemma~\ref{lemma-no-seq-spec}, 
we model any given task $T$ through a sequential object $S$ with two operations, {\Set}
and {\Get}, that each process access in a specific way: it first invokes {\Set} with its input to the task $T$ (receiving no output) and 
later invokes {\Get} in order to get an output value from $T$. Intuitively, decoupling the single operation of $T$ into two
(atomic) operations allows us to model concurrent behaviors that a single (atomic) operation cannot specify.
In what follows, let $SSpec(S)$ be the set with all sequential executions of $S$ 
in which each process invokes at most two operations, first {\Set} and then {\Get}, in that order.

\begin{theorem}
\label{theo-two-ops}
For every task $T = (\m{I},\m{O},\Delta)$ there is a sequential object $S$ with two operations, $\mathsf{set}(\id_i, x_i)$ and $\mathsf{get}(\id_i):y_i$, 
such that there is a bijection $\alpha$ between $VE(T)$ and $SSpec(S)$ satisfying that:
(1) each invocation or response of process $\id_i$ is mapped to an operation of process $\id_i$,
and (2) each invocation $inv$ (response $resp$) with input (output) $x$ is mapped to a completed {\Set}  ({\Get}) operation with input (output) $x$.
\end{theorem}

\begin{figure}[tb]
\centering{
\begin{minipage}[t]{150mm}
\footnotesize
\begin{tabbing}
aa\=aaaa\=aaaa\=aaaa\=\kill 

{\bf State:} a pair $(\sigma, \tau)$ of input/output simplexes, initialized to $(\emptyset, \emptyset)$\\ \\

{\bf Function} {\Set}($\id_i, x_i$)\\

\> {\bf Pre-condition:} $\id_i \in \ID \wedge \id_i \notin \ID(\sigma)$\\

\> {\bf Post-condition:} $\sigma' \leftarrow \sigma \cup \{ (\id_i, x_i) \}$\\

\> {\bf Output:} $\mathsf{void}$\\

{\bf endFunction}\\ \\

{\bf Function} {\Get}($\id_i$)\\

\> {\bf Pre-condition:} $\id_i \in \ID(\sigma) \wedge \id_i \notin \ID(\tau)$\\

\> {\bf Post-condition:} Let $y_i$ be any output value such that $\tau \cup \{ (\id_i, y_i) \} \in \Delta(\sigma)$. \\

\> \> \> \> \hspace{0.5cm} Then, $\tau' \leftarrow \tau \cup \{ (\id_i, y_i) \}$\\

\> {\bf Output:} $y_i$\\

{\bf endFunction}
\end{tabbing}
\footnotesize
\end{minipage}
\caption{A Generic Sequential Specification of a Task $T = (\m{I},\m{O},\Delta)$.}
\label{fig-generic-spec}
}
\end{figure}


An implication of Theorem~\ref{theo-two-ops} is that if one is analyzing an algorithm that uses a building-block (subroutine, algorithm, etc.) $B$ that solves a task $T$, 
one can safely replace $B$ with 
the sequential object $S$ related to $T$ described in the theorem 
(each invocation to the operation of $B$ is replaced with an (atomic) invocation to {\Set} and then an (atomic) invocation to {\Get}), 
and then analyze the algorithm considering the atomic operations of~$S$. The advantage of this transformation is that 
(1) if all operations in an algorithm are atomic, we can think that each process takes a step at a time in an execution, 
hence obtaining a a transition system with atomic events,
(2) at all times we have a concrete state of $S$ in an execution (which is not clear in a task specification) and
(3) given a state of $S$, an output for a {\Get} operation can be easily computed using the sequential object $S$ 
(something that is typically complicated for $B$ as it might be accessed concurrently).

The construction used (for simplicity) in the proof of Theorem~\ref{theo-two-ops} (in the full version of the paper) might be too coarse 
to be helpful for analyzing an algorithm. 
We would like to have a construction producing an equivalent sequential automaton modeling the task in a simpler way.
Consider the simple sequential object in Figure~\ref{fig-generic-spec} obtained from any given task $T = (\m{I},\m{O},\Delta)$, which is described
in a classic pre/post-condition form. Intuitively, the meaning of a state $(\sigma, \tau)$ is the following:
$\sigma$ contains the processes that have invoked the task so far
(this represents the \emph{participating set} of the current execution) while $\tau$ contains the outputs that have been produced so far.
The main invariant of the specification is that $\tau \in \Delta(\sigma)$. It directly follows from the properties of the task:
when a process invokes {\Set}$(\id_i, x_i)$, we have that $\tau \in \Delta(\sigma \cup \{(\id_i, x_i)\})$ 
because $\Delta(\sigma) \subset \Delta(\sigma \cup \{(\id_i, x_i)\})$, and when a process invokes {\Get}$(\id_i)$,
it holds that $\tau \cup \{(\id_i, y_i)\} \in \Delta(\sigma)$ because $\Delta(\sigma)$ is a pure complex of dimension $\dim(\sigma)$ 
and thus there must exist a simplex in $\Delta(\sigma)$ (properly) containing $\tau$ and with an output for $\id_i$.
One can formally prove that this sequential object and the one in the proof Theorem~\ref{theo-two-ops}
define the same set of sequential executions.

Finally, one can obtain ad-hoc and equivalent specifications for specific tasks, 
like the one for splitters in Figure~\ref{fig-splitter-spec} in Section~\ref{sec:ma-renaming}.

\section{Related Work}

\paragraph{Linearizability Criteria.}
Neiger observed for the first time that some fundamental tasks, 
like \emph{set agreement}~\cite{C93} and \emph{immediate snapshot}~\cite{BG93},
cannot be modeled as sequential objects~\cite{N94} (with a single operation).
Motivated by the need of a unified framework for tasks and objects, 
he proposed \emph{set-linearizability}~\cite{N94}.  
Roughly speaking, a set sequential object is generalization of a sequential object in which
transitions between states involve more than one operation (formally, a set of operations),
meaning that these operations are allowed to occur concurrently, and their results can be \emph{concurrency-dependent}. 
Set linearizability is the consistency condition for set-sequential objects, where one needs to find 
linearizability points (same as in linearizability) and several operations can be linearized at the same point
(different from linearizability). 

Later on, it was again observed that for some concurrent objects it is
impossible to provide a sequential specification, and
\emph{concurrency-aware} linearizability was defined~\cite{HRV15}.  
Set linearizability and
concurrency-aware linearizability are very closely related, both based on the same principle:
sets of operations can occur concurrently.  
Also, a non-automatic verification technique for
reasoning about concurrency-aware objects is presented in~\cite{HRV15}. 

Recently it was observed in~\cite{CRR18} that some natural tasks specify concurrency dependencies 
that are beyond the set-linearizability and concurrency-aware formalisms, hence that paper proposed \emph{interval linearizability}.
In an interval-sequen\-tial object not only sets of operations can occur concurrently but some of these 
operations might be pending and then overlap operations in the next transition; thus each operation
corresponds to an interval instead of a single point. Interval linearizability is the related consistency condition
in which, for each operation, one needs to find an interval in which the operation happens.
It is shown in~\cite{CRR18} that interval-linearizability is \emph{complete} for tasks in the sense 
that it is possible to specify \emph{any} task as an interval-sequential object (with a single operation). 

Although interval-sequential specifications can model any task, this approach does not seem to be
useful when one is searching for machine-checked proofs of concurrent algorithms. 
The main reason is that by replacing a task with its equivalent interval-sequential object,
we obtain a transition system in which one still needs to think in concurrent behaviors,
which is usually hard to deal with. 
In contrast, our proposed get-set transformation allows to ``decouple'' the inherent concurrency in tasks
in a way that in the resulting transition system all events are atomic, namely, they happen one after the other.

\paragraph{Mechanized Verification of Distributed Algorithms.}

Mechanized (or machine-assisted) verification of distributed and concurrent algorithms is usually done with model checking or theorem proving or a combination of both. Enumerative model-checking is the oldest fully automatic method with tools like Spin~\cite{SPIN2004} or TLC, the \TLA model checker~\cite{Lamport2002}. To avoid the well-known problem of state explosion, various optimisations such as symmetry or reduction have been introduced, and recent work is on going on parameterized model checking, for instance with MCMT (Model Checking Modulo Theory)~\cite{GhilardiR10}, Cubicle~\cite{ConchonGKMZ12} or ByMC~\cite{DBLP:conf/fmcad/JohnKSVW13}. Nevertheless, automatic verification of a distributed/concurrent algorithm is still restricted to small finite instances of the algorithm or imposes significant constraints on its description, due to the limited expressiveness of the specification language.

Fully automatic theorem proving is based on a proof decision procedure. For useful logics, it is often semi-decidable at best and heavily depends on heuristics to achieve good performance. Recent work on SMT has made a substantial leap forward checking complex formulae combining first-order reasoning with decision procedures for theory such as arithmetic, equality, arrays. Nonetheless, the overall proof of a distributed algorithm is still largely manual and, when seeking confidence in this proof, an interactive proof assistant is the current approach. Several examples of verification of complex distributed algorithms exist: Chord with Alloy~\cite{Zave2012}, Pastry with \TLA~\cite{Lu2011,Lu2013}, Paxos also with \TLA~\cite{Lamport2011}, snapshot algorithms in Event-B~\cite{Andriamiarina2014}, just to cite a few.

Several wait-free implementations of tasks have been mechanically proven (e.g.~\cite{OHearnRVYY10,TofanSR14,DragoiGH13}). However, to the best of our knowledge, no non-trivial algorithm built upon concurrent tasks have been mechanically proved. Our intuition for this situation is that proofs cannot be made modular and compositional when using bricks which are inherently concurrent if their internal structure must be visible to take into account this concurrency. Several complex and original algorithms can be found in the literature such as Moir and Anderson renaming algorithm~\cite{MoirA95} that we have considered in this paper, stacks implemented with elimination trees~\cite{ST97}, lock-free queues with elimination~\cite{MoirSSS05}. In these papers, the correctness proofs are intricate as they must consider the algorithm as a whole, including the tricky part involving wait-free objects, and they have not been mechanically checked. Our approach which exposes a more simple and sequential specification (instead of a complex concurrent implementation) seeks to alleviate this limitation.

\section{Final Remarks and Future Work}

In this paper, we showed a technique to circumvent the known impossibility of specifying a task as a sequential object.
Our technique consists in modeling the single operation of the task with two atomic operations, {\Set} and {\Get}.
This transformation leads to a framework for developing transitional models of concurrent algorithms using 
tasks and sequential objects as building blocks. As a proof of concept, we developed in a companion paper~\cite{HQ19} a full and modular 
\TLA proof of the Moir\&Anderson renaming algorithm~\cite{MoirA95}.

A natural extension of our work is to apply the framework to other concurrent algorithms.
Another direction is to extend our techniques to the case of \emph{refined tasks} and 
\emph{interval-sequential} objects, recently defined in~\cite{CRR18}.
These two formalisms are generalization of the task and sequential object formalism with 
strictly more expressiveness; particularly, contrary to the task formalism, refined task are \emph{multi-shot},
namely, each process may perform several invocations, possibly infinitely many.

A third direction is to study if the duality between 
the epistemic logic approach and the topological approach shown in~\cite{GLR18} 
might be useful in verifying concurrent algorithms.
Generally speaking, it is shown in~\cite{GLR18} that a task can be represented as a \emph{Kripke model}
with an \emph{action model}, specifying the knowledge obtained by  processes when solving the task.
It could be interesting to explore how this knowledge could be reflected in our {\Set}/{\Get} construction 
and if it could be useful in proving correctness.

\subsubsection*{Acknowledgements.}
Armando Casta\~{n}eda was supported by PAPIIT project IA102417.

\bibliography{proofCheck-arXiv}{}
\bibliographystyle{plain}


\newpage
\appendix

\section{Proofs and Extra Material of Section~\ref{sec-dealing-tasks}}
\label{app-tasks}

\subsection{Model of Computation}

We consider a standard concurrent system
with $n$ \emph{asynchronous} processes, $p_1, \hdots, p_n$,
which may \emph{crash} at any time during an execution of the system, i.e., 
stopping taking steps (for more detail see for example~\cite{HS08,R13}).
{Processes communicate with each other by invoking operations
{on} shared, concurrent \emph{base objects}.
A base object can provide $\R/\W$ operations (also called \emph{register}), 
more powerful operations, such as  $\TAS, \FAA, \SWAP \textrm{~or~} \CAS$, 
or solve a concurrent distributed problem, for example, $\mathsf{Splitter}$, $\mathsf{Renaming}$ or $\mathsf{Set\_Agreement}$.

Each process follows a local state machines $A_1, \hdots, A_n$,
where $A_i$ specifies which operations on base objects $p_i$ executes
in order to return a response when it invokes a high-level operation (e.g. $\mathsf{push}$ or $\mathsf{pop}$ operations).
Each of these base-objects operation invocations is a \emph{step}.
An \emph{execution} is a possibly infinite sequence of steps
and invocations and responses of high-level operations,
with the following properties:
\begin{enumerate}

\item Each process first invokes a high-level operation, and only when it has a corresponding response,
it can invoke another high-level operation, i.e., executions are \emph{well-formed}.

\item For any invocation $inv(\langle \mathsf{opType}, p_i, input \rangle)$ of a process $p_i$,
the steps of $p_i$ between that invocation and {its corresponding response (if there is one),
are steps} that are specified by $A_i$ when $p_i$
invokes the high-level operation $\langle \mathsf{opType}, p_i, input \rangle$.
\end{enumerate}

A high-level operation in an execution is \emph{complete} if
both its invocation and response appear in the execution.
An operation is \emph{pending}
if only its invocation appears in the execution.
{A process is \emph{correct} in an execution if it takes infinitely many steps.}

\subsection{Sequential Specifications}

A central paradigm for specifying distributed problems is that of a shared object that processes may access
concurrently~\cite{HS08,R13}, but the object is defined in terms
of a sequential specification, i.e., an automaton describing the
outputs the object produces when it is accessed sequentially.

A \emph{sequential object} $X$ is specified by a (not necessarily
finite and possibly non-deterministic) Mealy state machine
$(Q,{Inv},{Res},\delta)$, where $Inv$ is the set with all possible invocations to the object and $Res$ is the set with all possible responses from the object.
The responses are determined both by
its current state $s\in Q$ and the current input $in \in {Inv}$.
If $X$ is in state $q$ and it receives as input an invocation $in\in Inv$ by
process $p$, then, if $(q',r) \in \delta(q,in)$, the meaning is that
$X$ may return the response $r$ to the invocation $in$ by process
$p$, and move to state $q'$.  Notice that
there may be several possible responses
(if the object is non-deterministic), however, it is convenient to
assume that the next state $q'$ is uniquely determined by the response $r$,
namely, if $(q',r), (q'',r) \in \delta(q,in)$, then we have $q' = q''$. 
Also, it is convenient to require that the object $X$ is
\emph{total}, meaning that for any state $q$, $\delta(q,in)\neq
\emptyset$, for all $in\in {Inv}$.

For any sequence of invocations $in_0,\ldots,in_m$,
a \emph{sequential execution of $X$} starting in $q_0$ is
 $$q_0,in_0,r_0,q_1,in_1,r_1,\ldots,q_m,in_m,r_m$$ where $q_0$ is an
initial state of $X$, and $(q_{i+1},in_{i+1}) \in \delta(q_i,in_i)$.
However, given that we require that the object's response at a state
uniquely determines the new state, we may denote the execution by
 $$in_0,r_0,in_1,r_1,\ldots,in_m,r_m,$$ because the sequence of states
$q_1,\ldots,q_m$ is uniquely determined by $q_0$, and by the sequences
of invocations and responses. Without loss of generality we only consider sequential automata with a single initial state for each object.

The \emph{sequential specification} of an object $X$,  $SSpec(X)$, 
is the set of all its sequential executions. Notice that 
$SSpec(X)$ is \emph{prefix-closed}: if an execution is in $SSpec(X)$,
so is the execution obtained by removing the last invocation and its response.

Figure~\ref{fig-spec-TestAndSet} presents a sequential specification of the well-known
$\mathsf{Test\&Set}$ object, which has been used in a large number of concurrent algorithms (see for example~\cite{HS08,R13});
the specification is presented in the usual pre/post-condition specification style.
Intuitively, the object is initialized to 0 and the first invocation 
obtains 0 (the winner) and the rest obtain 1 (the losers).

\begin{figure}[htb]
\centering{
\begin{minipage}[t]{150mm}
\footnotesize
\begin{tabbing}
aa\=aaaa\=aaaa\=aaaa\=\kill 

{\bf State:} Integer $X$ initialized to $0$\\ \\

{\bf Function} {\sf Test\&Set}()\\

\> {\bf Pre-condition:} none\\ 

\> {\bf Post-condition:} \\ 

\> \> $temp \leftarrow X$\\

\> \> $X' \leftarrow 1$\\

\> {\bf Output:} \\ 

\> \> $temp$\\

{\bf endFunction}

\end{tabbing}
\footnotesize
\end{minipage}
\caption{Sequential Specification of $\mathsf{Test\&Set}$.}
\label{fig-spec-TestAndSet}
}
\end{figure}

Once we have a sequential specification, there are various ways of defining what it means
for an execution to \emph{satisfy} an object, namely, that it respects the sequential
specification. \emph{Linearizability}~\cite{HW90} is the standard notion used to identify
correct executions of implementations of sequential objects.
Intuitively, an execution is linearizable if its operations can be
ordered sequentially, without reordering {non-overlapping} operations,
so that their responses satisfy the specification of the implemented object.
To formalize this notion we define} a partial order on
the completed operations of an execution $E$:
${\sf op} <_E \mathsf{op}'$ if and only if $res({\sf op})$ precedes
$inv(\mathsf{op}')$ in $E$.
Two operations are \emph{concurrent} if they are incomparable by $<_E$.
The execution is \emph{sequential} if $<_E$ is a total order.

\begin{definition}
An execution $E$ is \emph{linearizable} with respect to $X$ if there is
a sequential execution $S$ of $X$
(i.e., $S \in SSpec(X)$) such that
\begin{enumerate}

\item
$S$ contains every completed operation of $E$
and might contain some pending operations.
Inputs and outputs of invocations and responses in $S$ agree with inputs and outputs in~$E$.

\item
For every two completed operations $\mathsf{op}$ and $\mathsf{op}'$ in $E$,
if $\mathsf{op} <_E \mathsf{op}'$, then $\mathsf{op}$ appears before $\mathsf{op}'$ in $S$.
\end{enumerate}
\end{definition}

Using the linearizability correctness criteria for sequential objects we can define the set of \emph{valid} executions for $X$, denoted $VE(X)$.
Arguably, the set $VE(X)$ contains the behavior one might expect from a \emph{building-block} (e.g. an algorithm) that implements $X$
(i.e. all its executions are linearizable w.r.t. X).
\[
VE(X) = \big\{ E | E \hbox{ has only invocations and responses and is linearizable w.r.t. } X \big\}
\]

\subsection{Tasks}

\subsubsection{Definition of a Task}

A task is the basic distributed equivalent of a function in sequential computing, defined by a
set of inputs to the processes and for each (distributed) input to the
processes, a set of legal (distributed) outputs of the processes,
e.g.,~\cite{HKRbook}.  In an algorithm designed to solve a task, each
process starts with a private input value and has to eventually decide
irrevocably on an output value. 
A process $p_i$ is initially not
aware of the inputs of other processes.  Consider an execution where
only a subset of $k \leq n$ processes participate; the others crash without
taking any steps. A  set of pairs
$\sigma=\{(\id_1,x_1),\dots,(\id_k,x_k)\}$ is used to denote the input values,
or output values, in the execution, where  $x_i$ 
denotes the value of the process with identity $\id_i$, either an input
value or an output value. 

A set $\sigma$ as above is called a \emph{simplex}, and if the values
are input values, it is an \emph{input simplex}, if they are output
values, it is an \emph{output simplex}.  The elements of $\sigma$ are
called \emph{vertices}. An \emph{input vertex} $v=(\id_i,x_i)$
represents the initial state of process $\id_i$, while an \emph{output
  vertex} represents its decision.  The \emph{dimension} of a simplex
$\sigma$, denoted $\dim(\sigma)$, is $|\sigma|-1$, and it is
\emph{full} if it contains $n$ vertices, one for each process.  A
subset of a simplex, which is a simplex as well, is called a
\emph{face}.  Since any number of processes may crash, simplexes of
all dimensions are of interest, for taking into account executions
where only processes in the simplex participate. Therefore, the set of
possible input simplexes forms a \emph{complex} because its sets are
closed under containment. Similarly, the set of possible output
simplexes also form a complex.

More generally, a \emph{complex} $\m{K}$ is made of a set of vertices
$V(\m{K})$, and a set of simplexes (i.e. a set of sets), each simplex
being a finite, nonempty subsets of $V(\m{K})$, satisfying: (1) if
$v\in V(\m{K})$ then $\{v\}$ is a simplex of $\m{K}$, and (2) if
$\sigma$ is a simplex of $\m{K}$, so is every nonempty subset of
$\sigma$.  The dimension of $\m{K}$ is the largest dimension of its
simplexes, and $\m{K}$ is \emph{pure} of dimension $k$ if each of its
simplexes is a face of a $k$-dimensional simplex. In distributed
computing, the simplexes (and complexes) are often \emph{chromatic},
since each vertex $v$ of a simplex is labeled with a distinct process
identity.  The set of processes identities in an input or output
simplex $\sigma$ is denoted $\ID(\sigma)$.

\begin{definition}[Task]
A \emph{task} $T$ for $n$ processes is a triple $(\m{I},\m{O},\Delta)$
where $\m{I}$ and $\m{O}$ are pure chromatic $(n-1)$-dimensional
complexes, and $\Delta$ maps each simplex $\sigma$ from $\m{I}$ to a
subcomplex $\Delta(\sigma)$ of $\m{O}$, satisfying:
\begin{enumerate}
\item $\Delta(\sigma)$ is pure of dimension $\dim(\sigma)$,
\item For every $\tau$ in $\Delta(\sigma)$ of dimension
  $\dim(\sigma)$, $\ID(\tau) = \ID(\sigma)$,
\item If $\sigma,\sigma'$ are two simplexes in $\m{I}$ with
  $\sigma'\subset \sigma$ then $\Delta(\sigma')\subset\Delta(\sigma)$.
\end{enumerate}
\end{definition}

A task has only one operation, let us call it $\mathsf{task}()$, which
process $\id_i$ may call with value $x_i$ only if $(\id_i,x_i)$ is a
vertex of $\m{I}$.  The operation ${\sf task}(x_i)$ may return $y_i$
to the process only if $(\id_i,y_i)$ is a vertex of $\m{O}$.  A task is a
very compact way of specifying a distributed problem, and indeed
typically it is hard to understand what exactly is the problem being
specified.  Intuitively, $\Delta$ specifies, for every simplex
$\sigma\in\m{I}$, the valid outputs $\Delta(\sigma)$ for the processes
in $\ID(\sigma)$ assuming they run to completion, and the other
processes crash initially, and do not take any steps.

As with other frameworks for specifying concurrent objects 
(e.g. linearizability for sequential specifications), tasks have their own correctness
criteria that defines the executions satisfying a given task. 
Let $E$ be an execution where each process invokes a task $\langle \m{I},
\m{O}, \Delta \rangle$ once.  Then, $\sigma_{E}$ is the input simplex
defined as follows: $(\id_i,x_i)$ is in $\sigma_{E}$ iff in $E$ there
is an invocation of ${\sf task}(x_i)$ by process $\id_i$.  The output
simplex $\tau_{E}$ is defined similarly: $(\id_i,y_i)$ is in
$\tau_{E}$ iff there is a response $y_i$ to a process $\id_i$ in $E$.
We say that $E$ \emph{satisfies} $(\m{I}, \m{O}, \Delta)$ if for
every prefix $E'$ of $E$, it holds that $\tau_{E'} \in
\Delta(\sigma_{E'})$. Note that it might be the case that $\dim(\tau_{E'}) \leq \dim(\sigma_{E'})$.

The prefix requirement prevents executions that globally seem correct,
but in a prefix a process predicts future invocations. This requirement
has been implicitly considered in the past by stating that an
algorithm solves a task if any of its executions agree with the task
specification.

Using the satisfiability notion of tasks we can now consider the set of valid executions, $VE(T)$,
for a given task $T = (\m{I},\m{O},\Delta)$.
Arguably, the set $VE(T)$ contains the behavior one might expect from a \emph{building-block} (e.g. an algorithm) that implements $T$.
\[
VE(T) = \big\{ E | E \hbox{ has only invocations and responses and satisfies } T \big\}
\]

\subsubsection{The Splitter Task}

As an example consider the \emph{splitter} task~\cite{MoirA95} defined informally as follows. Each process invokes ${\sf splitter}$ with its $\ID$ as input 
and outputs {\Stop}, {\Down} or {\Right}. For every $0 < k \leq n$, it is required that if $k$ processes invoke the splitter (note necessarily concurrently),
at most one process outputs {\Stop}, at most $k-1$ output {\Down} and at most $k-1$ output {\Right}.
Formally, the splitter task $T_\mathsf{spl} = (\m{I}_\mathsf{spl}, \m{O}_\mathsf{spl}, \Delta_\mathsf{spl})$ is defined as:

\begin{enumerate}

\item The vertices of the input complex $\m{I}_\mathsf{spl}$ are all pairs of the form $(\id_i, \id_i)$, for every ID process $\id_i$.

\item $\m{I}_\mathsf{spl}$ is the complex made of the $(n-1)$-dimensional simplex $\{(\id_1,\id_1),\dots,(\id_n, \id_n)\}$ (and all its faces),  
with all distinct $\ID$ processes $\id_1, \hdots, \id_n$.

\item The vertices of the output complex $\m{O}_\mathsf{spl}$ are all pairs of the form $(\id_i, \mathsf{stop})$, $(\id_i, \mathsf{down})$ and $(\id_i, \mathsf{right})$ for every ID process $\id_i$.

\item Given a simplex $\tau = \{(\id_1,y_1),\dots,(\id_m, y_m)\}$ with vertices in $\m{O}_\mathsf{spl}$ and an integer $k$,
let $SP(\tau, k)$ be the \emph{splitter predicate} that holds only if 
\begin{enumerate}
\item all $\id_i$s are distinct,
\item $|Stop| \leq 1$, $|Down| \leq k-1$ and $|Right| \leq k-1$,
where $Stop = \{ \id_i | y_i = \mathsf{stop} \}$, $Down = \{ \id_i | y_i = \mathsf{down} \}$ and $Right = \{ \id_i | y_i = \mathsf{right} \}$.
\end{enumerate}

\item $\m{O}_\mathsf{spl}$ contains every $(n-1)$-dimensional simplex $\tau$ (and all its faces), 
such that $SP(\tau, n)$ holds.

\item For every $(k-1)$-dimensional input simplex $\sigma$,
 $\Delta_\mathsf{spl}(\sigma)$ contains every $(k-1)$-dimensional output simplex $\tau$ (and all its faces) such that
$\ID(\tau) = \ID(\sigma)$ and $SP(\tau, k)$ holds.
\end{enumerate}

Figure~\ref{fig-splitter-app} shows a graphic description of the splitter task for three processes with IDs 1, 2 and 3.
The input complex, shown at the left, consists of a triangle and all its faces. The output complex, at the right, 
contains all possible valid output simplexes (the triangle with all {\Right} outputs is not in the complex).
The $\Delta$ function maps the input vertex with $\ID$ $1$ to the output vertex $(1,\mathsf{stop})$, the input edge 
with $\ID$s 1 and 2 to the complex with the bold edges in the output complex and the input triangle is mapped to
the whole output complex. The rest of $\Delta$ is defined symmetrically.

\begin{figure}[tb]
\centerline{\includegraphics[width=9cm]{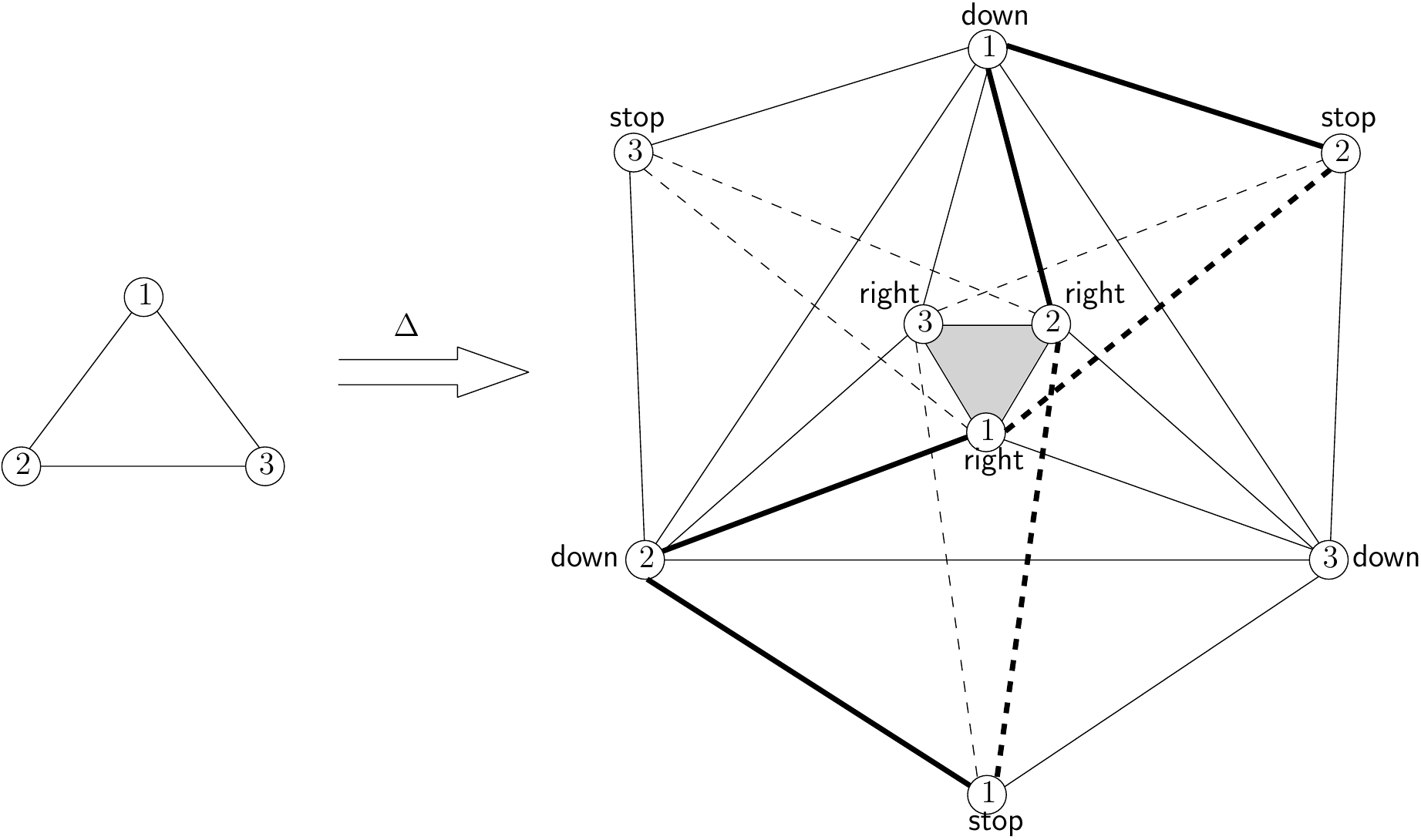}}
\caption{(repeated) The Splitter Task for Three Processes.}
\label{fig-splitter-app}
\end{figure}

\subsubsection{The Exchanger Task}

A second interesting example is the Java \emph{exchanger} object which is informally defined as follows in the Java documentation: 
\begin{quotation}
\it A synchronization point at which threads can pair and swap elements within pairs. 
Each thread presents some object on entry to the exchange method, matches with a partner thread, and receives its partner's object on return.
\end{quotation}

Clearly the object is informally specified in terms of concurrent executions, very much in the style of the task formalism.

Exchangers have been used in~\cite{HSY10} to implement a concurrent stack, and the lack of a sequential specification of exchangers
makes the proof in that paper intricate. They have also been used in a number of concurrent implementations, 
e.g.~\cite{SLS09,ST97}. More precisely,  in \cite{ST97}, Shavit and Touitou present the implementation of pools and stacks with \emph{elimination trees}, a form of \emph{diffracting trees}~\cite{ShavitZ96} which achieves high efficiency at high contention levels. A simplified version of their algorithm is the following. There are two kinds of opposite requests: enqueue and dequeue for a stack. The structure is constructed from \emph{elimination balancers} that are connected to one another to form a balanced binary tree. Each leaf of the tree holds a standard concurrent stack implementation (e.g. with locks). Each internal node of the tree holds a \emph{prism} and an \emph{exchanger}. The prism has an internal state (0 or 1) and two outputs labelled 0 and 1. It routes a request according to this state: an enqueue request goes on the output labelled as the internal state, a dequeue request goes on the output labelled as the inverse of the internal state. The internal state is flipped after each request. This allows the requests to spread on the tree while ensuring that a dequeue follows the same path as the most recent enqueue. To speed things up and to avoid contention of the internal state, two mechanisms are added. First, two concurrent requests of the same kind are directly routed on both output without changing the internal state. Secondly, an \emph{exchanger} is used to pair opposite requests: when both an enqueue and a dequeue are present, they are matched, they swap their values and they directly exit the tree without being further propagated (observe that this version of the exchanger is slightly different than the one above as 
processes exchange opposite requests). The actual implementation uses an array of prisms to avoid the bottleneck of the root and first-levels balancers, however this does not change the overall specification of the algorithm.

Although there is no sequential specification of exchanger in the literature (a proof such as the one for lemma~\ref{lemma-no-seq-spec} shows that there does not exist such a specification), one can succinctly define it as a task. Intuitively, in order processes exchange values, 
an exchanger matches pairs of processes, with the possibility that some processes are unmatched (marked as matched with a default value denoted $\bot$).
The exchanger task $T_\mathsf{exc} = (\m{I}_\mathsf{exc}, \m{O}_\mathsf{exc}, \Delta_\mathsf{exc})$ is defined as follows.

\begin{enumerate}

\item The vertices of the input complex $\m{I}_\mathsf{exc}$ are all pairs of the form $(\id_i, \id_i)$, for every ID process $\id_i$.

\item $\m{I}_\mathsf{exc}$ is the complex made of the $n$-dimensional simplex $\{(\id_1,\id_1),\dots,(\id_n, \id_n)\}$ (and all its faces),  
with all distinct $\ID$ processes $\id_1, \hdots, \id_n$.

\item The vertices of the output complex $\m{O}_\mathsf{exc}$ are all pairs $(\id_i, \id_j)$ and $(\id_i, \bot)$, where $\id_i$ and $\id_j$ are distinct process IDs.

\item Given a simplex $\tau = \{(\id_1,y_1),\dots,(\id_m, y_m)\}$ with vertices in $\m{O}_\mathsf{exc}$,
let $EX(\tau)$ be the \emph{exchanger predicate} that holds only if 
\begin{enumerate}
\item all $\id_i$'s are distinct,
\item $\id_i$ is matched with a different process or not matched at all: 
$y_i \in \{\id_1, \hdots, \widehat \id_i, \hdots, \id_m, \bot \}$, where circumflex ($\, \widehat \, \,$) denotes omission, 
\item $\id_i$ is matched with at most one process, namely, it appears in a second entry at most once,
\item matches are consistent, i.e., if $y_i = \id_j$ then $y_j = \id_i$.
\end{enumerate}

\item $\m{O}_\mathsf{exc}$ contains every $n$-dimensional simplex $\tau = \{(\id_1,y_1),\dots,(\id_n, y_n)\}$ (and all its faces)
such that $EX(\tau)$ holds.

\item For every $(k-1)$-dimensional input simplex $\sigma$,
 $\Delta_\mathsf{exc}(\sigma)$ contains every $(k-1)$-dimensional output simplex $\tau$ (and all its faces) such that 
$\ID(\tau) = \ID(\sigma)$ and $EX(\tau)$ holds.
\end{enumerate}

\begin{figure}[tb]
\centerline{\includegraphics[width=11cm]{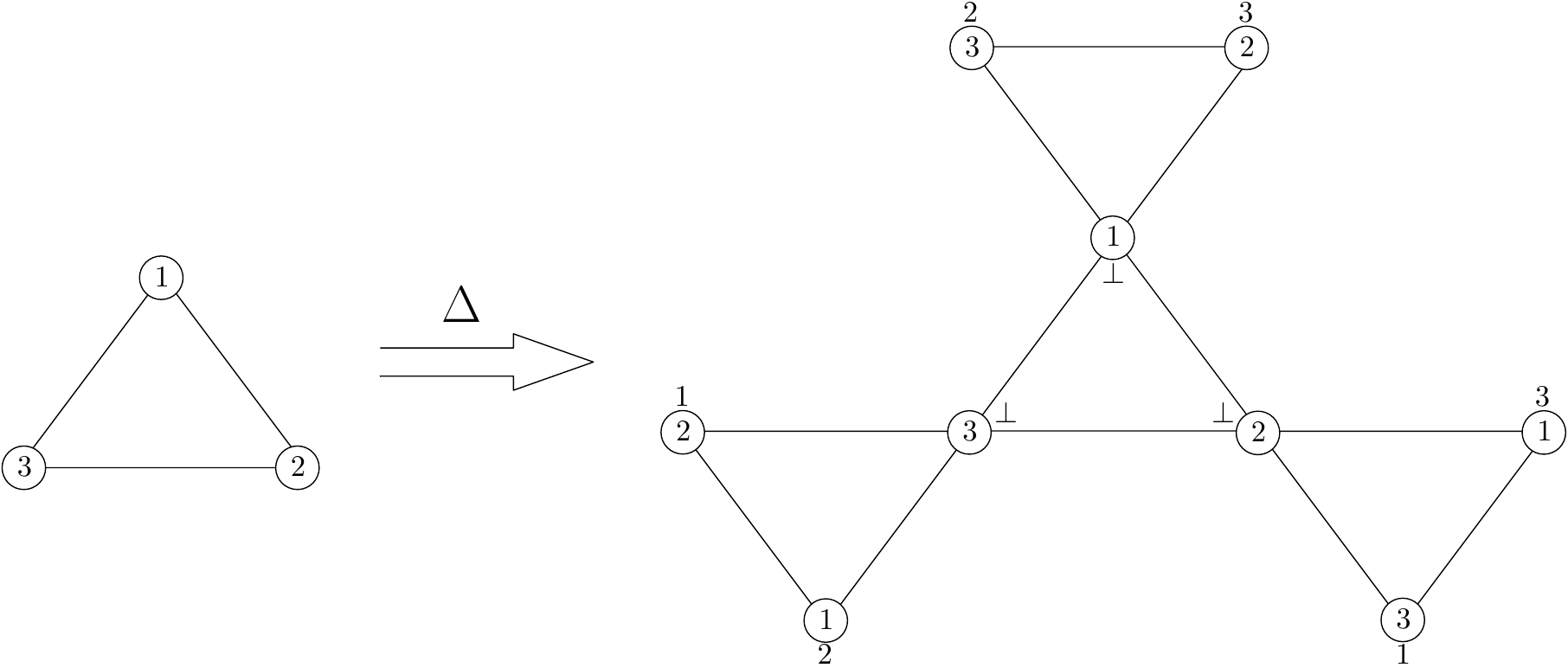}}
\caption{The Exchanger Task for Three Processes.}
\label{fig-exchanger}
\end{figure}

The exchanger task for three processes with $\ID$s 1, 2 and 3 is depicted in Figure~\ref{fig-exchanger}.
$\Delta$ maps the input vertex $i$ to $(i, \bot)$ and the edge with $\ID$s $i$ and $j$ to the complex
with edges $\{ (i,\bot), (j, \bot) \}$ and $\{ (i,j), (j, i) \}$, and the input triangle is mapped to the whole output complex.


\subsection{Modeling Tasks as Sequential Objects}
\label{sec-task-sequential}

Intuitively, tasks and sequential specifications are inherently different paradigms for specifying distributed problems:
while a task specifies what a set of processes might output when running concurrently, a sequential specification
specifies the behavior of a concurrent object when accessed sequential (and linearizability tells when a 
concurrent execution ``behaves'' like a sequential execution of the object).

A natural question is if any task can be modeled as a sequential object with a single operation, namely, the object defines the same set of valid executions. 
A well-known example for which this is 
possible is the consensus distributed coordination problem that can be equivalently defined as a task or as a sequential 
object (see for example~\cite{HS08} where it is defined as an object\footnote{Sometimes the object is defined with two operations 
(in the style of the Theorem~\ref{theo-two-ops}), however, consensus can be equivalently defined with one operation.} 
and~\cite{HKRbook} where it is defined as a task). 
Another interesting example is the $\mathsf{Test\&Set}$ atomic operation that is typically specified through a sequential object,
however it can also be specified as a task. 
Figure~\ref{fig-test-and-set} depicts the $\mathsf{Test\&Set}$ task for three processes 
(the specification in Figure~\ref{fig-spec-TestAndSet} is not one-shot but it can be easily made one-shot by adding that restriction in the pre-condition).
In general, this is not the case, as the following result shows.\\

\begin{figure}[tb]
\centerline{\includegraphics[width=9cm]{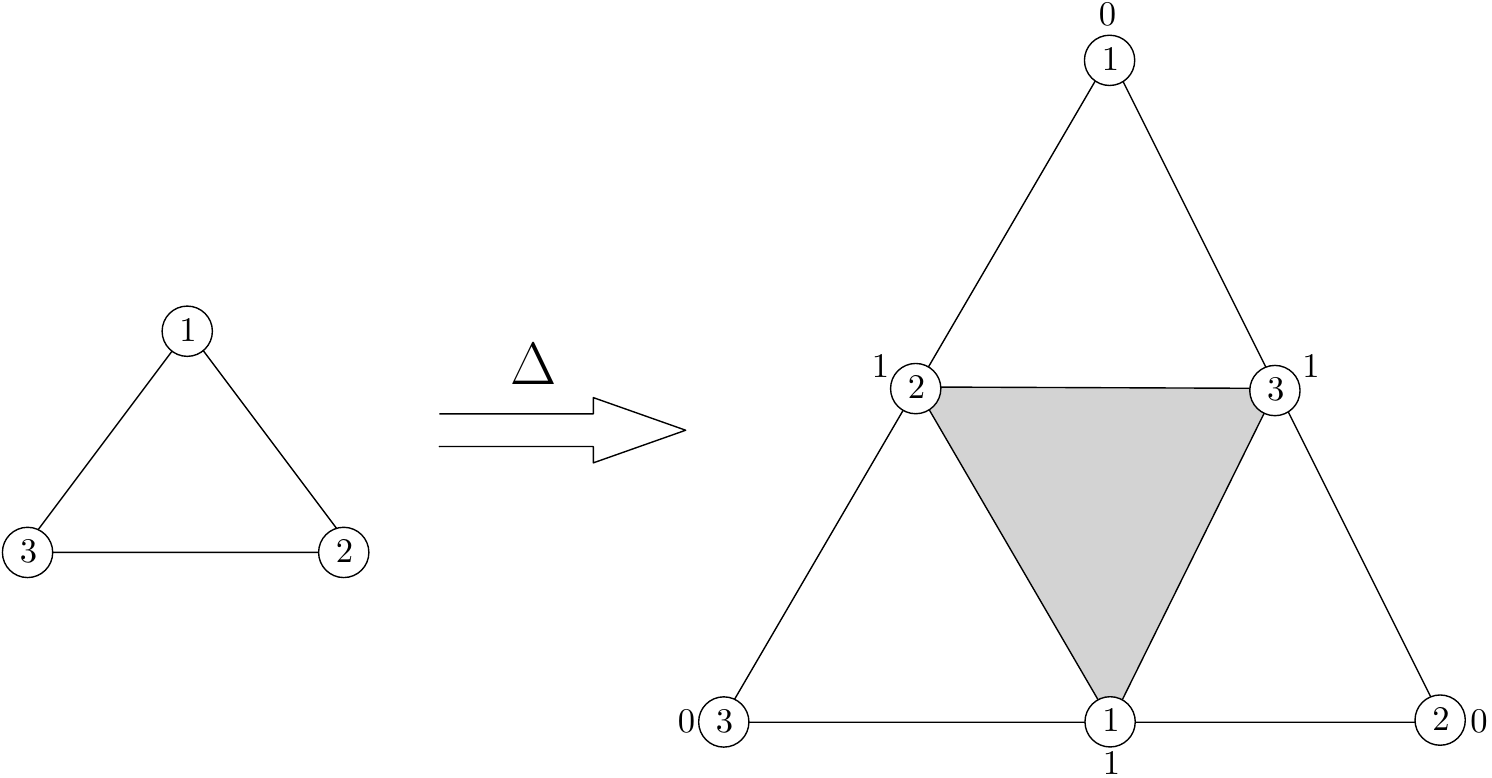}}
\caption{The one-shot $\mathsf{Test\&Set}$ Object for Three Processes Modeled as Task.}
\label{fig-test-and-set}
\end{figure}


\noindent
{\bf Lemma~\ref{lemma-no-seq-spec} (repeated)}. Consider the splitter task $T_\mathsf{spl} = (\m{I}_\mathsf{spl},\m{O}_\mathsf{spl},\Delta_\mathsf{spl})$.
There is no sequential object $X_\mathsf{spl}$ with a single operation satisfying:
\[
VE(T_\mathsf{spl}) = VE(X_\mathsf{spl}).
\]

\begin{proof}
Suppose by contradiction that there is such an object $X_\mathsf{spl}$ and 
consider the following fully concurrent execution for three processes:
\[\begin{array}{ll}
E = & inv(p_1, p_1) ; inv(p_2, p_2) ; inv(p_3, p_3) ; resp(p_1):\mathsf{down} ; resp(p_2):\mathsf{down} ;\\ & resp(p_3):\mathsf{right}. 
\end{array}\]
For a prefix $E'$ of $E$, one can  verify that $\tau_{E'} \in \Delta_\mathsf{spl}(\sigma_{E'})$;
for example, $\sigma_E = \{(p_1, p_1), (p_2, p_2), (p_3, p_3)\}$, $\tau_E = \{(p_1,\mathsf{down}), (p_2, \mathsf{down}), (p_3, \mathsf{right})\}$ 
and $\tau_E \in \Delta_\mathsf{spl}(\sigma_E)$.
Then, $E$ satisfies $T_\mathsf{spl}$, from which follows that $E \in VE(T_\mathsf{spl})$.

Now, our assumption implies that $E \in VE(X_\mathsf{spl})$, thus $E$ is linearizable with respect to $X_\mathsf{spl}$.
Without loss of generality suppose that there is a linearization $S$ of $E$ in which $inv(p_1, p_1) ; resp(p_1):\mathsf{down}$ is the first linearized operation. 
Thus, $S$ is a sequential execution of $X_\mathsf{spl}$, namely, $S \in SSpec(X_\mathsf{spl})$.
Since $SSpec(X_\mathsf{spl})$ is prefix-closed and $F = inv(p_1, p_1) ; resp(p_1):\mathsf{down}$ is a prefix of $S$,
we have that $F \in SSpec(X_\mathsf{spl})$. 
This is a contradiction because $F$ is indeed an execution which is linearizable with respect to $X_\mathsf{spl}$ ($F$ is a linearization of itself),
hence $F \in VE(X_\mathsf{spl})$, but F does not satisfy $T_\mathsf{spl}$ (clearly $\tau_F \notin \Delta_\mathsf{spl}(\sigma_F)$), 
and thus $F \notin VE(T_\mathsf{spl})$, which is a contradiction.
\end{proof}

In a very similar way one can prove that the exchanger task defined above and 
the following known tasks cannot be specified as sequential objects
with a single operation:

\begin{enumerate}

\item Adaptive renaming~\cite{ABDPR90}. Processes start with distinct inputs names taken from the space $[1, \hdots, N]$ and decide 
distinct outputs names from the space $[1, \hdots, M]$, with $N >> M$. It is required that if $k \leq n$ processes run concurrently, 
the output names belong to $[1, \hdots, f(k)]$, for some function $f : {1, \hdots, n} \rightarrow \{1, \hdots, N\}$,
i.e., the output space is on function on the number of participating processes.

\item Set agreement~\cite{C93}. It is a generalization of the
well-known consensus where processes propose values and have to agree on at most $k$ proposals.

\item Immediate snapshot~\cite{BG93}. It is a task
  which plays an important role
 in distributed computability~\cite{HKRbook}.  A process can write a
 value to the shared memory using this operation, and gets back a
 snapshot of the shared memory, such that the snapshot occurs
 immediately after the write.

\item Adopt-commit~\cite{BGLR,G98} is a concurrent object which 
  proved to be is useful to simulate round-based
protocols for set-agreement and consensus.  Given an input $u$ to the
object, the result is an output of the form $(commit,v)$ or
$(adopt,v)$, where $commit/adopt$ is a decision that indicates whether
the process should decide value $v$ immediately or adopt it as its
preferred value in later rounds of the protocol.

\item Conflict detection~\cite{AE14} is a task that has been shown to be equivalent to the 
{adopt-commit}. Roughly, if at least two different values are proposed
concurrently at least one process outputs true.
\end{enumerate}

To circumvent the impossibility result in the previous lemma, 
we model any given task $T$ through a sequential object $S$ with two operations, {\Set}
and {\Get}, that each process access in a specific way: it first invokes set with its input to the task $T$ (receiving no output) and 
later invokes {\Get} in order to get an output value from $T$. Intuitively, decoupling the single operation of $T$ into two
(atomic) operations allows us to model concurrent behaviors that a single (atomic) operation cannot specify.
In what follows, let $SSpec(S)$ be the set with all sequential executions of $S$ 
in which each process invokes at most two operations, first {\Set} and then {\Get}, in that order.\\


\noindent
{\bf Theorem \ref{theo-two-ops} (repeated)}. For every task $T = (\m{I},\m{O},\Delta)$ there is a sequential object $S$ with two operations, {\Set}$(\id_i, x_i)$ and {\Get}$(\id_i):y_i$, 
such that there is a bijection $\alpha$ between $VE(T)$ and $SSpec(S)$ satisfying that
\begin{enumerate}
\item each invocation or response of process $\id_i$ is mapped to an operation of process $\id_i$,
\item each invocation $inv$ (response $resp$) with input (output) $x$ is mapped to a completed {\Set} ({\Get}) operation with input (output) $x$.
\end{enumerate}

\begin{proof}
We define $S$ as follows. The sets of invocation and responses, $Inv$ and $Res$, of $S$ 
contain $inv(\mathsf{set}, \id_i, x_i)$ and $res(\mathsf{set}, \id_i, x_i):{\sf void}$, respectively,
for each input vertex $(\id_i, x_i) \in \m{I}$. 
Similarly, for each output vertex $(\id_i, y_i) \in \m{O}$,
$Inv$ and $Res$ contain $inv(\mathsf{get}, \id_i)$ and $res(\mathsf{get}, \id_i):y_i$.

For every execution $E \in VE(T)$, $S$ has a state $s_E$
and the initial state of $S$ is $s_\xi$, where $\xi$ denotes the empty string.
We define the transition function $\delta$ of $S$ inductively as:
\begin{enumerate} 

\item For every execution $E \in VE(T)$ consisting of only one invocation $inv(\id_i, x_i)$ (i.e. $E = inv(\id_i, x_i)$), 
we define 
\[ \delta(s_\xi, inv(\mathsf{set}, \id_i, x_i)) = \{ (s_E, res(\mathsf{set}, \id_i, x_i):{\sf void})\}. \]

\item For every execution $E \in VE(T)$ with the form $E = E' \cdot e$, for some non-empty $E'$ prefix,
$\delta$ is defined as:

\begin{enumerate}

\item If $e= inv(\id_i, x_i)$, then 
\[ \delta(s_{E'}, inv(\mathsf{set}, \id_i, x_i)) = \{ (s_{E' \cdot e}, res(\mathsf{set}, \id_i, x_i):{\sf void}) \}. \]

\item If $e = res(\id_i):y_i$, then  
\[ \delta(s_{E'}, inv(\mathsf{get}, \id_i)) = \{(s_{E' \cdot e}, res(\mathsf{get}, \id_i):y_i)\}. \]
\end{enumerate}
\end{enumerate}

Observe that $S$ is a deterministic automaton whose sequential executions are precisely the executions in $VE(T)$
(one can think that $S$ is an automaton that recognizes the language $VE(T)$). Moreover, each invocation $(\id_i, x_i)$ in an execution in 
$VE(T)$ induces a transition in $S$ with an invocation to {\Set}$(\id_i, x_i)$ and, similarly, each response $(\id_i, y_i)$ in an execution in 
$VE(T)$ induces a transition in $S$ with an invocation to {\Get}$(\id_i)$ whose response value is $y_i$.
Thus, the desired bijection $\alpha$ in $VE(T) \rightarrow SSeq(S)$ is precisely obtained from the definition of $S$.
\end{proof}

\begin{figure}[tb]
\centering{
\begin{minipage}[t]{150mm}
\footnotesize
\begin{tabbing}
aa\=aaaa\=aaaa\=aaaa\=\kill 

{\bf State:} a pair $(\sigma, \tau)$ of input/output simplexes, initialized to $(\emptyset, \emptyset)$\\ \\

{\bf Function} {\Set}($\id_i, x_i$)\\

\> {\bf Pre-condition:} \\

\> \> $\id_i \in \ID \wedge \id_i \notin \ID(\sigma)$\\

\> {\bf Post-condition:} \\

\> \> $\sigma' \leftarrow \sigma \cup \{ (\id_i, x_i) \}$\\

\> {\bf Output:} \\

\> \> $\sf void$\\

{\bf endFunction}\\ \\

{\bf Function} {\Get}($\id_i$)\\

\> {\bf Pre-condition:} \\

\> \> $\id_i \in \ID \wedge \id_i \notin \ID(\tau)$\\

\> {\bf Post-condition:} \\

\> \> Let $y_i$ be any output value such that $\tau \cup \{ (\id_i, y_i) \} \in \Delta(\sigma)$\\
\> \> $\tau' \leftarrow \tau \cup \{ (\id_i, y_i) \}$\\

\> {\bf Output:} \\ 

\> \> $y_i$\\

{\bf endFunction}
\end{tabbing}
\footnotesize
\end{minipage}
\caption{(repeated) A Generic Sequential Specification of a Task $T = (\m{I},\m{O},\Delta)$.}
\label{fig-generic-spec-app}
}
\end{figure}

An implication of Theorem~\ref{theo-two-ops} is that if one is analyzing an algorithm that uses a building-block (subroutine, algorithm, etc.) $B$ that solves a task $T$, 
one can safely replace $B$ with 
the sequential object $S$ related to $T$ described in the theorem 
(each invocation to the operation $B$ is replaced with an (atomic) invocation to {\Set} and then an (atomic) invocation to {\Get}), 
and then analyze the algorithm considering the atomic operations of~$S$. The advantage of this transformation is that 
(1) if all operations in an algorithm are atomic, we can think that each process takes a step at a time in an execution, 
hence obtaining a transition system with atomic events,
(2) at all times we have a concrete state of $S$ in an execution (which is not clear in a task specification) and
(3) given a state of $S$, an output for a {\Get} operation can be easily computed using the sequential object $S$ 
(something that is typically complicated for $B$ as it might be accessed concurrently).
In light of this, the construction used (for simplicity) in the proof of Theorem~\ref{theo-two-ops} might be too coarse to be helpful for analyzing an algorithm. 
Thus, we would like to have a
construction producing an equivalent sequential automaton modeling the task in a simpler way.

Consider the sequential object in Figure~\ref{fig-generic-spec} obtained from any given task $T = (\m{I},\m{O},\Delta)$, which is described
in a classic pre/post-condition form. Intuitively, the meaning of a state $(\sigma, \tau)$ is the following:
$\sigma$ contains the processes that have invoked the task so far
(this represents the \emph{participating set} of the current execution) while $\tau$ contains the outputs that have been produced so far.
The main invariant of the specification is that $\tau \in \Delta(\sigma)$. It directly follows from the properties of the task:
when a process invokes {\Set}$(\id_i, x_i)$, we have that $\tau \in \Delta(\sigma \cup \{(\id_i, x_i)\})$ 
because $\Delta(\sigma) \subset \Delta(\sigma \cup \{(\id_i, x_i)\})$, and when a process invokes {\Get}$(\id_i)$,
it holds that $\tau \cup \{(\id_i, y_i)\} \in \Delta(\sigma)$ because $\Delta(\sigma)$ is pure of dimension $\dim(\sigma)$ 
and thus there must exist a simplex in $\Delta(\sigma)$ (properly) containing $\tau$ and with an output for $\id_i$.
One can formally prove that this sequential object and the one in the proof Theorem~\ref{theo-two-ops} 
define the same set of sequential executions.

The formal definition of the sequential object in Figure~\ref{fig-generic-spec}  is the following.

\begin{enumerate}

\item For every $\sigma \in \m{I}$, and for every $\tau \in \m{O}$, 
$q_{(\sigma, \tau)}$ is a state in $Q$. The initial state is $q_{(\emptyset, \emptyset)}$.

\item For every input vertex $(\id_i, x_i) \in \m{I}$, $inv(\mathsf{set}, \id_i, x_i) \in Inv$ and $res(\mathsf{set}, \id_i, x_i):{\sf void} \in Res$.

\item For each output vertex $(\id_i, y_i) \in \m{O}$, $inv(\mathsf{get}, \id_i) \in Inv$ and $res(\mathsf{get}, \id_i):y_i \in Res$.

\item For every state $q_{(\sigma, \tau)}$, 
\begin{enumerate}

\item for every $(\id_i, x_i)$ such that $\id_i \notin \ID(\sigma)$ and $\sigma \cup \{(id_i,x_i)\} \in \m{I}$,
	\[ \delta(q_{(\sigma, \tau)}, inv(\mathsf{set}, \id_i, x_i)) = \{ (q_{(\sigma \cup \{(id_i,x_i)\}, \tau)}, res(\mathsf{set}, \id_i, x_i):{\sf void}) \}, \]

\item for every $(\id_i, y_i)$ such that $\id_i \in \ID(\sigma)$, $\id_i \notin \ID(\tau)$ and \mbox{$\tau \cup \{(id_i,y_i)\} \in \Delta(\sigma)$,}
	\[  (q_{(\sigma, \tau \cup \{(id_i,y_i)\})}, res(\mathsf{get}, \id_i): y_i) \in \delta(q_{(\sigma, \tau)}, inv(\mathsf{get}, \id_i)). \]
\end{enumerate}
\end{enumerate}

Finally, one can obtain simpler and equivalent specifications for specific tasks, like we did for the splitter in Section~\ref{sec:ma-renaming}.
Figure~\ref{fig-splitter-spec-app} presents such a specification where 
$\sigma$ is represented with the set $Participants$,
$\tau$ with the sets $Stop, Down$ and $Right$ and
the splitter predicate in the task is literally expressed in the get operation.
An ad hoc sequential specification of the exchanger is depicted in Figure~\ref{fig-exchanger-spec}
(a slight variation gives the exchanger used in~\cite{ST97}).

\begin{figure}[tb]
\centering{
\begin{minipage}[t]{150mm}
\footnotesize
\begin{tabbing}
aa\=aaaa\=aaaa\=aaaa\=\kill 

{\bf State:} Sets $Participants, Stop, Down, Right$, all initialized to $\emptyset$\\ \\

{\bf Function} {\Set}($\id$)\\

\> {\bf Pre-condition:}\\ 

\> \> $\id \notin Participants$\\

\> {\bf Post-condition:}\\ 

\> \> $Participants' \leftarrow Participants \cup \{ \id \}$\\

\> {\bf Output:} \\ 

\> \> $\sf void$\\

{\bf endFunction}\\ \\

{\bf Function} {\Get}($\id$)\\

\> {\bf Pre-condition:} \\ 

\> \> $\id \in Participants \wedge \id \notin Stop, Down, Right$\\

\> {\bf Post-condition:}\\ 

\> \> $D \leftarrow \emptyset$\\

\> \> {\bf if $|Stop| = 0$ then} $D \leftarrow D \cup \{\mathsf{stop}\}$\\ 

\> \> {\bf if $|Down| < |Participants|-1$ then} $D \leftarrow D \cup \{\mathsf{down}\}$\\ 

\> \> {\bf if $|Right| < |Participants|-1$ then} $D \leftarrow D \cup \{\mathsf{right}\}$\\ 

\> \> Let $dec$ be any value in $D$\\

\> \> {\bf if $dec = \mathsf{stop}$ then} $Stop \leftarrow Stop \cup \{\id\}$\\ 

\> \> {\bf if $dec = \mathsf{down}$ then} $Down \leftarrow Down \cup \{\id\}$\\ 

\> \> {\bf if $dec = \mathsf{right}$ then} $Right \leftarrow Right \cup \{\id\}$\\ 

\> {\bf Output:} \\ 

\> \> $dec$\\

{\bf endFunction}

\end{tabbing}
\footnotesize
\end{minipage}
\caption{(repeated) An \textit{ad hoc} Specification of the Splitter Task.}
\label{fig-splitter-spec-app}
}
\end{figure}

\paragraph{Correctness and Completeness.}
In the light of the ad hoc sequential specifications in Figures~\ref{fig-splitter-spec-app} and~\ref{fig-exchanger-spec},
consider the following question: how can we know if a 
given sequential specification $X$ with {\Get} and {\Set} operations corresponds to a task $T$, namely, it actually models $T$? 
That is to say, we consider the direction opposite to Theorem~\ref{theo-two-ops}, from sequential objects to tasks.
One way to obtain such a result is to show that there is an isomorphism between $X$ and the sequential automaton, say $S_T$, obtained
from the generic construction of Figure~\ref{fig-generic-spec-app}, instantiated with $T$.
A second equivalent approach is to verify that $X$ is \emph{correct}, i.e., it satisfies the input/output relation of $T$,
and \emph{complete}, namely, it specifies all possible executions in $VE(T)$.
Satisfying these two properties implies that $X$ and $S_T$ are isomorphic.

Formally, $X$ is \emph{correct w.r.t.$T$} if, for each of its executions $E \in SSpec(X)$,
$\tau_E \in \Delta(\sigma_E)$, where $\sigma_E$ is the simplex containing an invocation to $T$ for each (complete) {\Set} operation of $X$ in $E$, with same process and input value,
and, similarly, $\tau_E$ is the simplex containing a response from $T$ for each (complete) {\Get} operation of $X$ in $E$, with same process output value.

We say that $X$ is \emph{complete w.r.t.$T$} if for each execution of $E \in VE(T)$,
$S_E \in SSpec(X)$, where $S_E$ is the sequential execution obtained from $E$ by replacing each
invocation to $T$ in $E$ with a complete {\Set} operation of $X$, with same process and input value, 
and each response from $T$ in $E$ with a complete {\Get} operation of $X$, with same process and output value.


\begin{figure}[tb]
\centering{
\begin{minipage}[t]{150mm}
\footnotesize
\begin{tabbing}
aa\=aaaa\=aaaa\=aaaa\=aaaa\=\kill 

{\bf State:} Sets $Participants, Matching$, both initialized to $\emptyset$\\ \\

{\bf Function} {\Set}($\id$)\\

\> {\bf Pre-condition:} $\id \notin Participants$\\

\> {\bf Post-condition:} $Participants' \leftarrow Participants \cup \{ \id \}$\\

\> {\bf Output:} $\sf void$\\

{\bf endFunction}\\ \\

{\bf Function} {\Get}($\id$)\\

\> {\bf Pre-condition:} $\id \in Participants \wedge \{ \id, \cdot \} \notin Matching$\\

\> {\bf Post-condition:}\\ 

\> \> $Matched \leftarrow \{ \id^\ast | \{ \id^\ast, \cdot \} \in Matching \}$\\

\> \> $Free \leftarrow Participants \setminus Matched$\\

\> \> {\bf if $\id \in Matched$ then} \\

\> \> \> Let $\id^\ast$ be the value in Matched such that  $\{ \id, \id^\ast \} \in Matched$\\

\> \> {\bf else}\\ 

\> \> \> Let $\id^\ast$ be any value in $Free \cup \{ \bot \}$\\

\> \> \> $Matching' \leftarrow  Matching \cup \{ \{ \id, \id^\ast\} \}$\\

\> {\bf Output:} $\id^\ast$\\

{\bf endFunction}

\end{tabbing}
\footnotesize
\end{minipage}
\caption{An \textit{ad hoc} Specification of the Exchanger.}
\label{fig-exchanger-spec}
}
\end{figure}

\paragraph{On Adaptiveness.}

An interesting property of the splitter and ${\sf Test\&Set}$ sequential objects in Figures~\ref{fig-spec-TestAndSet} and~\ref{fig-splitter-spec} is that they do not take into
account the number of processes in the system, namely, the specification is the same for any number of processes.
This property is known as \emph{adaptiveness} and can be formalized as follows.

Consider an infinite set of processes $\ID = \{p_1, p_2, \hdots \}$.
Consider a distributed problem that is specified through an infinite family of sequential objects:
for every finite set $S \subset \ID$, let $X_S$ be a sequential object for processes in $S$.
The family of objects is \emph{adaptive} if for every two sets $S \subset S'$,
$SSpec(X_S) = SSpec(X_{S'}, S)$, where $SSpec(X_{S'}, S)$ is the subset of $SSpec(X_{S'})$
with operations of processes in $S$.

The notion of adaptiveness for tasks is defined similarly.
Consider a distributed problem that is specified through an infinite family of tasks:
for every finite set $S \subset \ID$, let $T_S = (\m{I}_S,\m{O}_S, \Delta_S)$ be a sequential object for processes in $S$.
The family of tasks is \emph{adaptive} if for every two sets $S \subset S'$,
$\m{I}_S \subset \m{I}_{S'}$ and for every $\sigma \in \m{I}_S$, $\Delta_S(\sigma) = \Delta_{S'}(\sigma)$.

\end{document}